\documentclass[prb,aps,twocolumn,showpacs,floatfix,eqsecnum]{revtex4}
\usepackage{graphicx}

\def\REM#1 {{[\bf #1 ]}}
\def\NOTE#1 {{}}
\def\rhostar {{\rho^*}}
\def\sumocc{{\sum_{q {\ \rm occ}}}}
\def\sumodd{{\sum_{r {\ \rm odd}=1}^{R-1}}}
\def\uy{b} 

\newcommand{\RR} {{\bf R}}
\newcommand{\SSS} {{\cal S}}
\newcommand{\kk} {{\bf k}}
\newcommand{\muLC} {{\mu^{\rm LC}}}

\begin{document}

\title{
Stripes and holes in a two-dimensional model of 
spinless fermions and hardcore bosons
}

\author{N. G.~Zhang and C.~L.~Henley}

\affiliation{Department of Physics, 
Cornell University, Ithaca New York 14853-2501}



\begin{abstract}
We consider a Hubbard-like model of strongly-interacting
spinless fermions and hardcore bosons 
on a square lattice, such that nearest neighbor occupation is forbidden. 
Stripes (lines of holes across the lattice 
forming antiphase walls between ordered domains)
are a favorable way to dope this system below half-filling. 
The problem of a single stripe 
can be mapped to a spin-1/2 chain, which allows
understanding of its elementary excitations and
calculation of the stripe's effective mass for
transverse vibrations. 
Using Lanczos exact diagonalization, we investigate
the excitation gap and dispersion of a hole on
a stripe, and the interaction of two holes. 
We also study the interaction of two, three, and four stripes,
finding that they repel, and 
the interaction energy decays 
with stripe separation as if they are hardcore particles
moving in one (transverse) direction. 
To determine the stability of an array of stripes
against phase separation into particle-rich phase
and hole-rich liquid, we  evaluate the liquid's
equation of state, finding 
the stripe-array is not stable for bosons
but is possibly stable for fermions.
\end{abstract}

\maketitle

\section{introduction}
\label{sec-intro}

Stripes have been an area of active study
in the high-temperature superconductivity research.
They are modulations of charge and spin densities, 
and can be static or dynamic.
In 1995, Tranquada and coworkers\cite{Exp} observed the coexistence
of superconducting and stripe domain order for 
the material La$_{1.6-x}$Nd$_{0,4}$Sr$_x$CuO$_4$
at around doping $x=0.125$. The stripes are
one-dimensional objects on a two-dimensional plane,
and they have been called ``self-organized 
one dimensionality.''\cite{ZaanenScience}
Search for stripes in the Hubbard and $t-J$ models
has been intense and is still going strong
(see Ref.~\onlinecite{EmKi} for
a review of some experimental results and theoretical 
considerations).
 
>From the theory side, there
are basically two roads. One approach is using
microscopic models such as the Hubbard
and $t-J$ models to study numerically 
stripe formation and phase separation. 
DMRG appears to be the best tool in such studies
and has been applied to the $t-J$ model in 
Refs.~\onlinecite{dmrg1,dmrg2}. Another approach to stripes
is to use effective field theories, treating
stripes as macroscopic objects (macrosopic
in the sense of many-particle collective motion). 
For example in Ref.~\onlinecite{Zaanenstrings}, Zaanen
considered stripes as ``a gas of elastic quantum strings
in $2+1$ dimensions.'' The microscopic, numerical approach can be used to
understand the mechanism for stripe formation
and to derive macroscopic parameters such as
stripe-stripe interaction energy and stripe stiffness, 
but it is limited by computational power. The macroscopic
approach on the other hand takes these 
macroscopic parameters as an input and can
obtain results relevant to experiments,
but it can explain less about mechanism.
We will follow the microscopic route which enables us to
obtain macroscopic parameters from diagonalization
data. 

\subsection{The model}

In this paper, we study stripes in a strongly-interacting
model of spinless fermions and hardcore bosons on
the square latttice. The Hamiltonian is
	\begin{equation}
H=-t\sum_{\langle i,j \rangle}\left(
c^\dagger_i c_j + c^\dagger_j c_i\right)
+V\sum_{\langle i,j \rangle} \hat n_i \hat n_j,
	\label{eq-Ham}
	\end{equation}
with periodic boundary conditions.
We study both the spinless fermion and the hardcore
boson versions of the model. $c^\dagger_i$ and $c_i$ are 
the spinless fermion or hardcore boson creation and 
annihilation operators at site $i$, $\hat n_i=c^\dagger_i c_i$
the number operator, $t$ the nearest-neighbor hopping amplitude, 
and $V$ the nearest-neighbor interaction.
At the each site there can
be at the most one particle. Furthermore, we study
the strong-correlation limit of the model and take
$V=+\infty$, i.e., infinite nearest-neighbor repulsion.

The spinless fermion and hardcore boson model with infinite
nearest-neighbor repulsion involves a significant
reduction of the Hilbert space as compared to the
Hubbard model. The Hubbard model
on the $4\times 4$ lattice at half-filling, with 8
spin-up and 8 spin-down electrons has $1,310,242$
states in the largest matrix block,
after reduction by particle conservation,
translation, and the symmetries of the square.\cite{Lin}
In our model with infinite $V$,
after using particle conservation and translation symmetry
(but not point group symmetry), 
the largest matrix for the $7\times 7$ system 
has $1,906,532$ states (for 11 particles). 
We can therefore compute for 
all fillings the $7\times 7$ system whereas 
for the Hubbard model $4\times 4$ is basically the limit.
This also means that at certain limits we can
obtain results that are difficult to obtain
in the Hubbard model, for example, we 
can exactly diagonalize a system of two length-8 stripes 
on the $8\times 28$ lattice.

This is one of the two papers that we are publishing
to systematically study the phase diagram of the 
spinless fermion and hardcore boson model on the
square lattice with infinite nearest-neighbor repulsion.
In the other paper,~\cite{dilutepaper} we study the dilute limit of this 
model Eq.~(\ref{eq-Ham}), focusing on the problem of a few
particles, and that limit is dominated by two-body
interactions. In the present paper, we study the
near-half-filled limit of our model, where stripes are
natural objects. 

\begin{figure}[ht]
\centering
\includegraphics[width=0.3\linewidth]{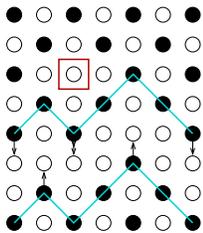}
\caption{
An isolated hole (shown in the box)
in a half-filled region cannot move.
A row of holes (a stripe) can slide along
in the vertical direction, with the arrows showing
the possible moves.
}
	\label{fig-hole}
	\end{figure}

With infinite nearest-neighbor repulsion,
the maximum filling fraction (particle per site)
for our model is 1/2, for both spinless
fermions and hardcore bosons. At exactly half-filling, the
particles form a checkerboard configuration that cannot move.
Adding a single hole to it cannot produce motion either 
because the hole is confined by neighboring particles.
A natural way to add holes to this system is to align
them in a row going across the system, as indicated
in Fig.~\ref{fig-hole}. We call this row
of holes a stripe and it is the subject of study in this
paper.

For many systems with certain aspect ratios (e.g., the 
$6\times 7$ lattice depicted in Fig.~\ref{fig-hole}),
the stripe state is the one with the 
smallest number of holes and nonzero energy.
Here we have an interesting kind of physics
of extended, fluctuating, quantum mechanical objects,
that are results of collective motions of many particles.

Spinless models related to the present one
have been invoked occasionally in the literature~\cite{Ko67}
but usually in the context of a specific question 
about spinfull models: it was recognized that
a spinless model would capture the same physics
with fewer complications.  The only systematic work 
on phase diagram of spinless fermions is by Uhrig.~\cite{uhrig}
However, the method is expansion around infinite
dimensionality; this will have special difficulty
with arrays of domain walls, since wall fluctuations and
the attendant kinetic energy are strongly suppressed in
high dimensions.  

The Falicov-Kimball model is an alternate way to simplify
the Hubbard model, which is also commonly presented in a
spinless form,~\cite{free93} and develops stripe-like incommensurate
patterns.~\cite{free02}  However, this model includes a second
immobile (classical) species of electron, so its stripes
cannot have quantum fluctuations.~\cite{Ba-KBiO3}

No known electron system realizes our model, even approximately. 
A ``half-metallic'' ferromagnet~\cite{halfmetal,Cu71}
-- meaning that for one spin
state, the conduction band is all filled or all empty -- 
realizes a spinless model for the other spin state. 
However in the best-known half-metals, the manganites,~\cite{Pa98} 
the formation of inhomogeneities is dominated by 
lattice distortions and orbital degeneracies (not to mention
static pinning disorder), so that quantum-fluctuating defects
find no role in current theories of manganites. 
(The same may be true for the Verwey transition in magnetite,~\cite{seo}
which was modeled previously with spinless electrons~\cite{Cu71}.) 
Conceivably the spinless model could be realized in an array of semiconductor 
quantum dots in a magnetic field.~\cite{manousakis}

The most plausible realization of the boson version of our model
would be an adsorbed gas of $^4$He on a substrate with 
square symmetry.  (The same model -- but on a triangular lattice --
has been introduced independently to model $^4$He 
on nanotube~\cite{green00,green02}). The fermion case could similarly
be realized, in principle, by spin-polarized $^3$He, however it would
be difficult to achieve a degree of polarization sufficient
to neglect the minority spin state.

This paper appears to be the first using exact diagonalization
(as opposed to DMRG)
to investigate the properties of 
interacting stripes with a microscopic Hamiltonian.~\cite{FN-EDstripe}

\subsection{Stripes in spinfull models}

In real fermion systems, stripes (antiphase domain walls 
containing holes) have been a prominent 
object of experimental study in the high-$T_c$ 
cuprates.~\cite{Exp,EmKi,Zaanenrev,Whiterev,Baskaran} 
Stripe-based mechanisms have been proposed for high-temperature
superconductivity,~\cite{EmKi-SC,EmKi} but
the prevalent current opinion is that stripes compete with
superconductivity.~\cite{Zaanenrev,Whiterev,Baskaran}

Nevertheless, 
stripes are obviously clues to how the charge and spin
degrees of freedom interact with each other, which is 
important in a majority of the high-$T_c$ theories. 
Stripes are modeled theoretically and numerically
using the same Hubbard model (or its variations) which 
were already accepted as models of homogeneous phases. 
It is still unsettled whether stripes are stabilized
in this model,  which omits long-range and even nearest-neighbor
Coulomb repulsion.  The most explicit calculations are by Density-Matrix
Renormalization Group (DMRG), adapted to two-dimensional systems formed
into strips.~\cite{dmrg1,dmrg2} The results favor stripes with 1/2 holes per
unit length, as found in experiments; however, those simulations are 
strongly influenced by the (necessary) open boundary conditions on two sides. 

What relation can our spinless model have to these 
spinfull models?
Firstly, at the intermediate scales appropriate to a stripe array, 
the spinfull and spinless systems are modeled in very similar ways:
stripes are mutually repelling, quantum-fluctuating strings. 
This level is appropriate to studies of the anisotropic 
transport expected in a (non-superconducting) stripe array, 
as well as the stripe interactions, as has been emphasized by
Zaanen.~\cite{zaanen,Eskes,Zaanenstrings}
In the spinfull models, only a little has
been done to explicitly relate the macroscopic parameters
to the underlying microscopic model.~\cite{smith}
In our model -- and unlike any spinfull model -- exact
diagonalization can address phenomena such as the
effective mass and interactions of carriers on a stripe,  
long-wavelength stripe fluctuations, stripe-stripe repulsion, 
or the transfer of carriers from one stripe to the next. 
Thus, not only analytically but numerically, one can go quite 
far in extracting such parameters for our model. 

Secondly, if spinfull stripes are stabilized at all
(in the absence of long-range repulsion), it is by
kinetic energy in the same fashion that our stripes are
stabilized.  In either case, a fermion
belonging to one domain hops transverse to the stripe, and 
(since the stripe is also an antiphase domain wall) still finds 
itself correctly placed in the order of the new domain. 
In our model, that order is the ``checkerboard'' pattern, 
which is enforced by the nearest-neighbor repulsion $V$. From 
this viewpoint, $V$ is analogous to the antiferromagnetic
coupling $J$ in the $t$-$J$ model. (Our $V\to \infty$ limit
then corresponds to the case $J \gg t$, which is the 
unphysical regime of the $t$-$J$ model.)

An even more precise correspondence would be to stripes in the
$t$-$J_z$ model:~\cite{Tch,CherNeto00} 
in that system (as in ours) the undoped, ordered state breaks 
a discrete (Ising-type) symmetry, 
and hence is practically inert, supporting
no gapless Goldstone (spinwave) excitations:
only the stripe itself has low-energy excitations. 
If the ordered state had a continuous symmetry, 
as in the Hubbard or $t$-$J$ models, 
spin-wave modes can mediate a $1/d^2$ attraction~\cite{Pryadko98}
between stripes (where $d$ is the stripe separation). 
If short-distance kinetic energy were to favor
a stripe array, the long-range force mediated by continuous spins
implies phase separation in the limit of 
very small doping.~\cite{FN-Casimir}

Our stripes with an occupation of 0.5 hole per unit length 
are insulating and correspond to insulating stripes of
1 hole per unit length in the $t$-$J$ model (the factor
of two reflects the density of the ordered background state), 
as implied in the 
original ideas~\cite{Pre93,Pre94,Zaanenrev} about stripe
stabilization from a strong-coupling viewpoint.

\subsection{Paper organization}

This paper is organized as follows.  First (in Sec.~\ref{sec-diagstripe}) 
we describe briefly our exact diagonalization
code for studying the near-half-filled limit of our model.
We use translation symmetry to block diagonalize
the Hamiltonian matrix. 
A graph viewpoint motivates a method of building the basis states
from a starting configuration and explains the
relation between boson and fermion energy spectra.

Next (Sec.~\ref{sec-onestripe})
we study the problem of a single stripe.
Here particles can only move in one direction, effectively reducing
the 2D problem to a 1D one. This one-stripe 
problem maps exactly to a spin-1/2 chain, which is
exactly solvable using Bethe ansatz.
Using the single-stripe energy dispersion relation along
the direction perpendicular to the stripe,
we obtain stripe effective mass for motion
in that direction.

In the next two sections, 
we study the problem of one or two holes on a stripe.
(For the one-hole case, the bosons and fermions have the same
energy spectrum.) 
We determine the dispersion relationship
(energy gap and effective mass) for one hole 
moving on a stripe, and study the binding of two holes. 
It is informative to study the finite-size dependence
of the energy on the lattice dimensions in both directions:
in particular, it decays exponentially as a function
of the lattice size perpendicular to the stripe, which
we explain quantitatively in terms of the stripe tunneling 
through a barrier between two potential wells. 

One of the motivations for studying our model
is our interest in a simple model of interacting 
stripes.  In Sec.~\ref{sec-twoandmore} we have exactly
diagonalized systems with two, three, and four stripes,
and find that the stripes repel.
The interaction energies scale as the inverse square
of the stripe spacing, 
which (like the above-mentioned tunneling)
follows from the one-dimensional nature of the
stripe's transverse motion in our systems. 


Finally (Sec.~\ref{sec-stripearray}) 
we discuss the stability of an array of stripes
by fitting the diagonalization
results in the intermediate filling limit
and using a Maxwell construction.
Our interest is whether at the intermediate-filling
limit we have the stripe-array case or a phase-separated
case with hole-rich regions and particle rich-regions.
The conclusion is, interestingly, that 
the boson stripe-array is not stable 
and the fermion stripe-array is very close to the
stability limit and is possibly stable.

Our earlier publication, Ref.~\onlinecite{HenleyZhang},
contains some of the results of this paper in a 
condensed form, with a focus on the stability of the stripe-array.
The present paper contains
more systematic and more updated results and a 
substantial number of 
important new results on, for example, 
boson and fermion statistics, stripe effective mass
and stripe-stripe interaction.

\section{Basis States and Diagonalization Program}
\label{sec-diagstripe}

In this section, we describe our exact diagonalization code. 
We begin by describing the conventions for labeling basis
states,  which are crucial to keep track of sign factors in
the fermion case.  Then we introduce a geometrical way to
view the basis states as nodes on a graph, and apply this
to classify the conditions under which the boson and fermion
problems have the same. 

\begin{figure}[ht]
	\centering
	\includegraphics[width=0.7\linewidth]{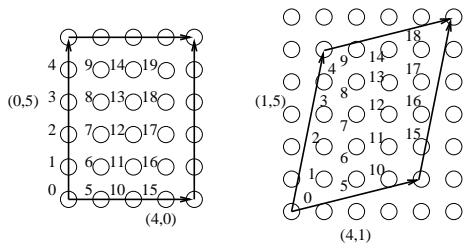}
    	\caption{Square lattices with periodic boundaries:
$(4,0)\times(0,5)$ on the left and $(4,1)\times(1,5)$ on the right.
Site numbers are shown, following the numbering convention,
upward and rightward.}
    	\label{fig-2lattice}
\end{figure}

\subsection{Basis states}
\label{sec-basis}

Each basis state with $M$ particles 
corresponds 1-to-1 to a {\it configuration}, 
which is array of the $M$ occupied site numbers 
$(i_1,i_2,...,i_M)$. 
Any configuration with two nearest neighbors occupied
is excluded because of the infinite $V$ in the Hamiltonian
Eq.~(\ref{eq-Ham}). 
Periodic boundary conditions are specified by two lattice vectors
${\bf R}_1$ and  ${\bf R}_2$ (which will be
called {\it boundary vectors} in the following), such that
for any lattice vector ${\bf r}$ we have
${\bf r}+n_1{\bf R}_1+n_2{\bf R}_2\equiv{\bf r}$,
where $n_1$ and $n_2$ are two integers.

The configuration by itself is insufficient to define the 
basis ket, since we need to specify its sign or phase factor. 
To fix the fermion sign, we must establish an arbitrary ordering 
of sites and always configurations  in this order. 
Then we denote the basis state 
$|n\rangle=c^\dagger_{i_1}c^\dagger_{i_2}...c^\dagger_{i_M}|0\rangle$, 

The site ordering convention used in this paper 
is to start with site 0 at the
lower left corner and move upward along a lattice column
until we encounter the cell boundaries defined by 
the boundary vectors ${\bf R}_1$ and ${\bf R}_2$, then 
shift rightward and repeating in the next column, progressing
until all sites have been numbered. 
Fig.~\ref{fig-2lattice} shows two example systems: $(4,0)\times(0,5)$
and $(4,1)\times(1,5)$.

In this basis the Hamiltonian Eq.~(\ref{eq-Ham}) acts as
	\begin{equation}
	H|n\rangle=(-t)\sum_{m\in {\cal M}} s(n,m) |m\rangle,
	\label{eq-Hn}
	\end{equation}
where ${\cal M}$ denotes the set of states 
created by hopping one particle in $|n\rangle$ 
to an allowed nearest-neighbor site.
For bosons $s(n,m)=1$ always and for fermions
$s(n,m)=\pm 1$. And the matrix
element is $\langle m | H|n\rangle = -s(n,m)t$ 
if $m\in {\cal M}$ and 0, otherwise.

We use lattice translation symmetry to block diagonalize
the Hamiltonian matrix $\langle m | H|n\rangle$.
The eigenstates that we use are the Bloch states\cite{Leung}
	\begin{equation}
	|n,{\bf k}\rangle=\frac{1}{N_{n{\bf k}}}\sum_{l=0}^{N-1} 
	e^{-i{\bf k}\cdot{\bf R}_l} T_l |n\rangle.
	\label{eq-Bloch}
	\end{equation}
In this expression ${\bf k}$ is a reciprocal lattice vector 
(one of $N$, where $N$ is the number of sites),
$\RR_l$ is a lattice vector 
(the order of $l$ in the sum is not important), 
$T_l$ is translation by $\RR_l$,
and $N_{n{\bf k}}$ is a normalization factor. 
The original basis states $|n\rangle$ are divided by translation
into classes and any two states in the same class
give the same Bloch state with an overall phase factor. 
A representative is chosen from each class
and is used consistently to build the Bloch states.
For a state $|n\rangle$ we denote its representative 
$|\bar n\rangle$, and the Hamiltonian matrix is block
diagonal in the sense that
$\langle \bar m, \kk'|H|\bar n,\kk\rangle=0$ when $\kk'\ne\kk$.
The matrix $\langle \bar m, \kk|H|\bar n,\kk\rangle$
is diagonalized for each $\kk$
vector using the well-known Lanczos method.

\subsection{State graph}
\label{sec-stategraph}

It is helpful conceptually to represent specific cases of
our Hamiltonian by the {\it state graph}, 
in an abstract (or high-dimensional)
space, such that each node corresponds to a basis state, 
and two nodes are joined by an edge of the graph
if and only if the two states differ by one particle hop. 
(As used in Sec.~\ref{sec-graphhop}, our many-body Hamiltonian
is almost equivalent to a single particle hopping on the
state graph.)
 The best-known precedent of this approach s Trugman's
study of one-hole and two-hole hopping in the $t$--$J_z$ model~\cite{Trugman}. 

Two states are {\it connected}
if one state can be changed to another by 
a succession of hops. 
For states near half filling, our $V=\infty$ constraint
forbids many hop moves, so the graph is sparse and can possess 
interesting topological properties relevant to our spectrum. 

\subsubsection{Droplets and accessibility of states}
\label{sec-droplets}

We remarked (Fig.~\ref{fig-hole})
that an isolated hole is unable to hop;
larger finite ``droplets'' of holes are still immobile, 
though they may gain kinetic energy from internal
fluctuations.  Specifically, \cite{Mila}
if one draws a rectangle with edges at $45^\circ$ to the
lattice around the droplet, so as to enclose every site which 
deviates from the checkerboard order, then (with the $V=\infty$ constraint)
sites outside this rectangle can never be affected by 
fluctuations of the droplet.

Due to the droplets, the state graph (representing
all states with $M$ particles) may
be broken up into many disconnected components, states
which cannot access each other by allowed hops. 
This happens if both boundary vectors are even vectors, so
that the system cell would support half filling with a perfect
checkerboard order.  For slightly less than 
half filling,  the {\it typical} basis state 
consists of scattered immobile droplets
of the kind just described.
Each component corresponds
to a different way to assign the holes to droplets, or merely
a shifted way of placing the same droplets. 
Each sort of droplet has its characteristic energy 
levels~\cite{FN-smalldroplets}
and the system energies are the sum, just like a system
of noninteracting atoms, each having its independent excited levels.

When the state graph is disconnected, the Hilbert space is 
correspondingly blocked into components which are not connected by 
matrix elements. This is beyond the 
blocking according to translational symmetry, which remains true. 

\subsubsection {Building basis set from a starting configuration}

In systems with even-even boundary conditions, 
we are not mainly interested in the configurations 
with droplets, but in those with two stripes.~\cite{FN-stripevsdroplet}
Since these are disconnected from the rest of the
Hilbert space, it speeds up the the diagonalization by a large factor 
if it is limited to the stripe-configuration 
subblock of the Hamiltonian matrix. 
The only way to separate out this subblock is to
generate basis states iteratively 
from a starting configuration which has two stripes. 

What of the case of even-odd or (equivalently) odd-odd boundary
conditions? These force {\it one} stripe (possibly with 
additional holes, or possibly more stripes). 
In every such case (excluding $L\times L$ with odd $L$), 
the stripe can fluctuate and absorb any droplets, so 
we believe the Hilbert space is fully connected, and
the diagonalization is not speeded up. 
Nevertheless, for every case near half-filling, 
we enumerated the basis from a starting configuration, 
as (we believe) this is the fastest way to do so. 
Systematically enumerating all possible fillings by brute force
would take more time than the diagonalization itself, since
almost every brute-force trial
configuration gets ruled out by the $V=\infty$ constraint.   

However, at low fillings (which we study in Ref.~\onlinecite{dilutepaper})
the brute-force enumeration is used. In that case, if we
generated from a starting configuration, there would be too many alternate
routes to reach a given state. Hence much time would be
wasted in searches to check whether each newly reached state was 
already on the basis state list. 

Below we outline the algorithm for building the basis states
and the Hamiltonian matrix.

	\begin{figure}[ht]
\centering
\includegraphics[width=\linewidth]{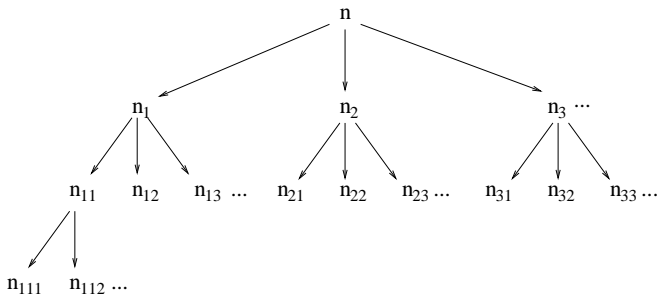}
\caption{
	Building basis states from a given state $n$. Applying
all allowed hoppings to $n$ (the level 0 state)
gives the states $n_1$, $n_2$, $n_3$, ...
Applying hoppings to $n_1$ (a level 1 state) 
gives level 2 states $n_{11}$, $n_{12}$, $n_{13}$, ...
The basis state list is then $n$, $n_1$, $n_2$, $n_3$, ...,
$n_{11}$, $n_{12}$, $n_{13}$, ..., $n_{21}$, $n_{22}$, ...,
$n_{111}$, $n_{112}$, ...
Note that when applying hoppings to a state, only add states
not already on the list. 
We call this graph of states connected by hopping
the {\it state graph}.
}
	\label{fig-tree}
	\end{figure}

To build basis states
without using translation symmetry from
a starting state $n$, we follow the 
following steps.

\begin{enumerate}

\item Apply hopping to $n$
and obtain states $n_1$, $n_2$, $n_3$,...
See Fig~\ref{fig-tree}. Form the basis
state list $n$, $n_1$, $n_2$, $n_3$,... that
are numbered 0, 1, 2, 3,... 
And record the Hamiltonian matrix elements 
$H_{0,1}$, $H_{0,2}$, $H_{0,3}$,...

\item Apply hoppings to the next state on
the basis state list that has not been applied hopping to.
Here from $n_1$ we get $n_{11}$, $n_{12}$, $n_{13}$...
Add these to the basis state list to form
$n$, $n_1$, $n_2$, $n_3$, ..., $n_{11}$, $n_{12}$, $n_{13}$, ...
(Note we should not add states that 
are already on the list. For example, applying hoppings to
$n_1$ will certainly give $n$ back again. Do not include this 
state in level 2 states. When computing Hermitian matrix elements
we only need $H_{p,q}$ for $p<q$.)
If $n_{11}$ is the $m$-th element on the list,
record the matrix elements $H_{1,m}$, $H_{1,m+1}$,
$H_{1,m+2}$,...

\item
Repeat the previous step until all
the states on the list have been applied hopping to and no
new states are created.

\end{enumerate}

We finish with a basis state list
$n$, $n_1$, $n_2$, $n_3$, ...,
$n_{11}$, $n_{12}$, $n_{13}$, ..., 
$n_{21}$, $n_{22}$, ..., $n_{111}$, $n_{112}$, ...
And we have stored the Hamiltonian matrix elements.

When building basis states with translation
symmetry, as in Eq.~(\ref{eq-Bloch}), the procedure is like the
above, except we apply hopping to
representative states of each class of
translation related states and also store
the representatives of the resulting states.

\subsection{Boson and fermion statistics}
\label{sec-statistics}

In this paper we study both the boson and the fermion
versions of our model Eq.~(\ref{eq-Ham}), and one question
that we often ask is when bosons and fermions
have the same spectrum. In this section, we
introduce a graph-based way to study the
relationship between the boson and 
fermion spectra.

Using Eq.~(\ref{eq-Hn}), we know that
if $\beta$ is a state that
can be obtained from 
the state $\alpha$
by one nearest-neighbor hop, 
then we have
	\begin{equation}
	\langle \beta | H | \alpha\rangle_f=s(\beta,\alpha)
	\langle \beta | H | \alpha\rangle_b,
	\label{eq-Hbf}
	\end{equation}
where the subscripts $f$ and $b$ denote the fermion and
boson matrix elements respectively and we have
used the fact that $s(\beta,\alpha)=1$
for bosons. If we can write 
	\begin{equation}
	s(\beta,\alpha)=\sigma(\beta) \sigma(\alpha),
	\label{eq-sab}
	\end{equation}
i.e., $s(\beta,\alpha)$ as a product of a function $\sigma$ that
depends on one state only, then Eq.~(\ref{eq-Hbf})
can be written as a matrix equation,
	\begin{equation}
	{\bf H}_f = {\bf \Sigma} {\bf H}_b {\bf \Sigma},
	\label{eq-matrixeq}
	\end{equation}
where $\Sigma_{\alpha,\beta}=\delta_{\alpha,\beta}\sigma(\alpha)$.
Because ${\bf \Sigma}={\bf \Sigma}^{-1}$, it is a similarity
transformation, and then the eigenvalues of ${\bf H}_f$ are identical to 
that of ${\bf H}_b$.  Indeed, this is a 
much stronger condition than having identical
eigenenergy spectra: the {\it eigenstates} are also
identical (modulo a sign), so e.g. an operator which 
depends on the basis state has the same expectation
in the boson and fermion cases. 
On the other hand, if the function $\sigma$
cannot be defined, the boson and fermion Hamiltonians are
not equivalent. 

\begin{figure}[ht]
	\centering
	\includegraphics[width=0.5\linewidth]{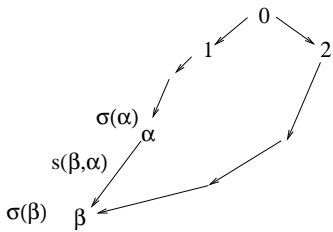}
	\caption{
	Obtaining $\sigma(\alpha)$, that is defined
	for each state, from a reference state
	0 and the step wise hopping sign function $s(\beta,\alpha)$,
	that is defined on the arrow pointing from $\alpha$
	to $\beta$. Define $\sigma(\beta)=s(\beta,\alpha)\sigma(\alpha)$.
	If the function $\sigma(\beta)$ is well-defined,
	i.e., all paths leading from 0 to $\beta$ 
	give the same sign, then the boson and fermion
	spectra are identical. 
	}
	\label{fig-sab}
	\end{figure}

We can relate Eq.~(\ref{eq-sab}) to the loops which occur in
virtually every state graph with the help 
of Fig.~\ref{fig-sab}. 
Here $s(\beta,\alpha)$ is defined 
on the arrow pointing from $\alpha$ to $\beta$, whereas
the function $\sigma(\alpha)$ is
defined on the node $\alpha$ of the graph.

Fig.~\ref{fig-sab} suggests a natural way to 
construct $\sigma(\alpha)$.  First we choose a reference point.
This should not matter and the one we choose is the starting
state of the exact diagonalization program, say $0$.
We set $\sigma(0)=1$. Then if one hop takes $\alpha$ 
to $\beta$, define $\sigma(\beta)=s(\beta,\alpha)\sigma(\alpha)$.
This is certainly correct if Eq.~(\ref{eq-sab}) is true; 
but if the state graph has a loop, two different paths lead to
the same state $\beta$ from 0, and
$\sigma(\beta)$ can be well defined only if they produce the same sign.
The answer is different for different state graphs (or disconnected 
subgraphs of the state graph), so it depends on the system dimensions
and filling. 

We can check numerically whether the function 
$\sigma(\alpha)$ is well defined
at the same time as we are building the basis set by applying hopping, 
i.e. constructing the state graph in the form of a tree
(Fig.~\ref{fig-tree}). 
We store a sign for each state,
starting with $\sigma(n)=1$ for state $n$
in Fig.~\ref{fig-tree}.  As we expand the
tree, we compute the sign for the next level of states.
Whenever we come to a state that is already on the list,  
we test whether the sign we produce following the current
path equals that already stored for that state.
Depending whether the sign always agrees or sometimes disagrees, 
we know the fermion and boson problems are or are not equivalent.
This method is the basis for statements later in the paper 
on boson and fermion spectra
(Sec.~\ref{sec-1stripeBoseFermi} and Sec.~\ref{sec-2andmoreBoseFermi}, 
and briefly in Secs.~\ref{sec-onehole} and \ref{sec-twoholes});
it is not necessary to compute all the eigenvalues. 

The boson-fermion equivalence condition can be re-stated
in a way that is independent of the site ordering convention
at the start of Sec.~\ref{sec-diagstripe}. 
Say each particle is originally numbered 
$1, \ldots, M$ and retains its number when it hops. 
Then $\sigma(\alpha,\beta)$ is simply the sign of the 
permutation which changes the string of particles
(when they are listed according to the site order). 
The function $\sigma(\alpha)$ is well-defined if and only
if the product of $\sigma(\alpha,\beta)$ around any closed
loop of the state graph is unity; in other words, 
if and only if any sequence of hops that restores the
original configuration induces an even permutation of the particles. 
When just two particles can be exchanged (an odd permutation), 
as is obviously possible at low fillings, the fermion and boson
spectra are obviously different.

\section{A Single Stripe}
\label{sec-onestripe}

\begin{figure}[ht]
\centering
\includegraphics[width=\linewidth]{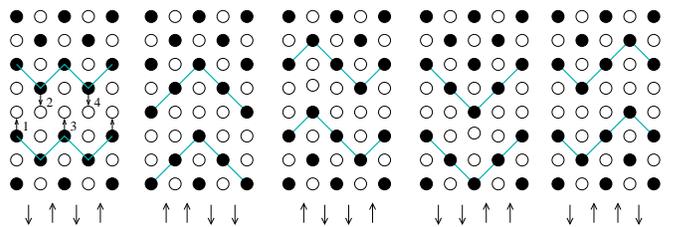}
\caption{
A single stripe in the $4\times 7$ system.
For the leftmost state, the four possible hops 
are labeled and the resulting configurations are shown
on the right. The corresponding spin chain 
configurations are shown below (see Sec.~\ref{sec-maptospin}). 
}
	\label{fig-stripe}
	\end{figure}

It was first observed by Mila, in a model very similar
to ours, that the creation of stripes is more favorable
than of isolated holes, as a way to dope the system below half
filling.~\cite{Mila,HenleyZhang}  We shall finally
address more rigorously the phase stability of the stripe 
(Sec.~\ref{sec-stripearray}), but
in this section we will study the eigenstates
of a single stripe in the system.
In Fig.~\ref{fig-stripe} we show a $L_x=4$ and $L_y=7$ 
system with 12 particles. A stripe of length four is formed
along the $x$ direction. 
The only four possible nearest-neighbor
hops of the leftmost state are indicated with arrows
and the resulting four states are shown on the right.
Note that with a single stripe, the particles can only
hop in the $y$ direction. It is not possible
for them to move in any fashion in the $x$ 
direction.

\begin{figure}[ht]
\centering
\includegraphics[width=0.6\linewidth]{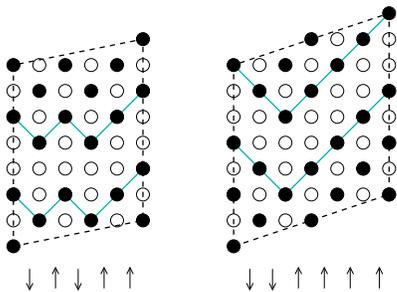}
\caption{
A single stripe in the $(5,1)\times(0,7)$
and the $(6,2)\times(0,7)$ systems.
The boundary conditions force a tilt of the stripe.
The corresponding spin-chains have
total $S_z$ equal to $1/2$ and $1$ respectively 
(see Sec~\ref{sec-maptospin}).
}
	\label{fig-tilted}
	\end{figure}

We also consider tilted boundaries
$(L_x,b)\times(0,L_y)$, which
force a tilt of $b$ steps to the stripe because
the site $(0,y)$ is identified with $(L_x,y+b)$.
In Fig.~\ref{fig-tilted}, the
$(5,1)\times(0,7)$ and $(6,2)\times(0,7)$
systems are shown. Note because each stripe step is at 45 degree 
angle, an even (odd) horizontal length of the stripe $L_x$
must produce an even (odd) number of total vertical steps $b$. 
Thus $L_x+b$ is always even. For a general
$(L_x,b)\times(0,L_y)$ system with $L_y$ odd,
the single-stripe states have $M=L_x(L_y-1)/2$ particles.
Note also that for the tilted stripes, vertical hopping is again
the only allowed motion of the particles.

Diagonal stripes
are observed in La$_2$NiO$_4$,\cite{Exp}
and they are an interesting topic
in the $t-J$ model.\cite{smith}
However, in our model with $V=\infty$, no hopping 
is possible away from a $45^\circ$ edge of any domain. 
Thus a diagonal stripe has no kinetic energy 
(unless there are additional holes, as
discussed in Sec.~\ref{sec-holekink})
and is disfavored. 

\subsection{Boson and fermion statistics for one stripe}
\label{sec-1stripeBoseFermi}
\label{sec-graphhop}


We find from diagonalization that for
one-stripe systems with rectangular boundaries,
the boson and fermion spectra are identical and 
are symmetric about zero. For
a stripe in systems with tilted boundaries, there
are complications. For the $(2m+1,1)\times(0,2l+1)$ system,
with $m$ and $l$ integers,
the spectrum is not symmetric about zero. In this case, 
with $l$ odd, the fermion spectrum is still identical 
to the boson spectrum, but for $l$ even, the fermion
energies are the boson ones with a minus sign (and because
the spectrum is no longer symmetric about zero, the fermion
energies are not the same as the boson energies).
For the $(2m,2)\times(0,2l+1)$ system, the boson and fermion spectra are
identical and symmetric about zero. 

All these numerical findings can be explained
in terms of the state graph, extending Sec.~\ref{sec-statistics}. 
We must characterize the permutations induced by a sequence of 
particle hops which return to the same basis configuration.
This is quite easy in the single stripe state, since 
(as already commented on Fig.~\ref{fig-stripe}
and Fig.~\ref{fig-tilted}),
particles move in the $y$ direction only. 
Our discussion includes all $(L_x, b) \times (0,L_y)$
systems (which are tilted boundaries if $b\neq 0$).
In these cases, particles are confined to the same column, 
and the net permutation
is a product of permutations in each column. 



Consider the total particle permutation induced 
when the system returns to the starting state 
by a loop on the state graph.~\cite{FN-signrule}
If in this process the stripe 
as a whole makes no net movement across the
$y$ boundaries, then in each column the particles 
undo their hops and the net permutation is the identity. 
Thus, the only nontrivial loops are those in
which the stripe crosses the $y$ boundary.
Indeed, since each hop moves the stripe locally by 
$\Delta y=\pm 2$, the stripe must cross
it must cross the boundary {\it twice}.
The particles in each column have returned to the
original positions with a cyclic permutation in each column, 
which has the sign 
   \begin{equation}
         \sigma_{\rm col} \equiv (-1)^{(L_y+1)/2}
   \label{eq-sigmacol}
   \end{equation}
Therefore the sign of the system's permutation  is
${\sigma_{\rm col}}^{L_x}$ which (since $L_x+b$ is even)
reduces to $(-1)^{b(L_y+1)/2}$: the fermion and boson spectra
differ if and only if this phase is $-1$, i.e.
only when $b$ is odd {\it and} $L_y\equiv 1 \pmod{4}$.

The $\pm E$ symmetry of the single-stripe spectrum
also falls out from visualizing the Hamiltonian
as a single particle hopping on the state graph
(with amplitude $-t = -1$ for every graph edge), and 
recalling the spectrum depends only on the attributes of closed loops.
We can always change our convention for the phase factors
of the basis states, which is equivalent to a ``gauge transformation''
on the nodes of the state graph. 
Whenever a state graph is {\it bipartite}, meaning
every closed loop on it has an even number of edges, 
it is well-defined to divide the basis states into
``even'' and ``odd'' classes. 
If we change the phase factor by $-1$ for every ``odd'' basis
state, every edge (every $s(n,m)$ factor) picks up a factor $-1$.
By gauge equivalence, the new Hamiltonian matrix has the 
same spectrum. Yet on the other hand, it is manifestly the same
as the old matrix with $t \to -t$, so it has the sign-reversed
spectrum; this proves the $\pm E$ symmetry. 
Now, the nontrivial loops in the state-graph are those
that pass the stripe twice across the $y$ boundary;
this motion requires $L_xL_y\equiv N$ hops. 
Thus, the state-graph is bipartite and the
spectrum has $\pm E$ symmetry, if and only if $N$ is even. 

If $N$ is odd, the gauge-invariant effect of reversing $t \to -t$ 
is to change the net product of $s(n,m)$ around 
every nontrivial loop by a factor $-1$.  
But we showed that switching Fermi and Bose statistics 
creates the same sign: hence the fermion spectrum is the
inverse of the boson spectrum, as observed. 

\subsection{Mapping to spin chain}
\label{sec-maptospin}

In Figs.~\ref{fig-stripe} and \ref{fig-tilted},
we have indicated a natural
map from any configuration $y(x)$ of the stripe to 
a state of a spin chain of length $L_x$. 
(Recall that ``configuration'' was defined in Sec.~\ref{sec-basis}
as a pattern of occupied sites; the map from the quantum 
states of the stripe is more subtle on account of phase factors, 
which are confronted later in this subsection.)
Here it is obvious the spin length is one-half, for  
each stripe step, $y(x+1)-y(x)$, 
can take only two values.\cite{FN-Tch}
The up step of the stripe is mapped to an up spin
and the down step to a down spin. 
For example, the leftmost configuration in Fig.~\ref{fig-stripe}
maps to a spin state $|\downarrow\uparrow\downarrow\uparrow\rangle$.
This mapping was first noted by Mila~\cite{Mila}, who used it
(as we do here) to evaluate the exact ground state energy of a stripe. 

Note that when the periodic boundary conditions are rectangular, 
the stripe satisfies $y(L_x)=y(0)$.
In the corresponding states of the spin chain,  
the total $z$ spin component is zero.
For tilted boundaries,
the map to the spin-chain works exactly as before,
the only difference being that the resulting
spin configurations have $\sum _i{S_{i}^z} = b/2$.

The stripe-to-spin-chain map is not one-to-one,
for translation in the $y$ direction of a single-stripe state gives
the same spin-chain state (with a possible fermion sign).
A rather analogous situation (but one dimension higher)
appears in the two-dimensional quantum dimer model 
(QDM).~\cite{rokhsar,henleydimer}
In the QDM, every dimer configuration corresponded with a
surface $z(x,y)$, uniquely except for an arbitrary shift of $z$, 
just as our spin (or 1D fermion) chain corresponds to a stripe $y(x)$ here. 
There were two kinds of low energy excitations of the QDM: 
ripplon modes of the ``surface'', and transverse motion (or tunneling)
of the ``surface'' through a periodic boundary condition. 
These correspond, respectively, to our ripplon modes labeled by
$k_x$ (see Sec.~\ref{sec-stripeexcite} and also Sec.~\ref{sec-onehole})
and to our $y$-translation modes labeled by $k_y$ 
(see Eq.~(\ref{eq-mexcite})).
The analogy is imperfect in that, in the QDM, the dimer configuration
is ``real'' and the surface is an abstract mapping of it, 
whereas here the spin chain is the abstract mapping and the stripe is ``real.''

\subsubsection {Spin Hamiltonian}

To determine the Hamiltonian in spin language, 
consider that each allowed particle hop takes
the stripe either from up-step/down-step to down-step/up-step
or from down-step/up-step to up-step/down-step. They correspond
to the nearest-neighbor spin flips 
$\uparrow\downarrow$ to $\downarrow\uparrow$ and 
$\downarrow\uparrow$ to $\uparrow\downarrow$ respectively. 

Since the matrix element of $H$ is $-t$ between states differing by a hop, 
this must also be the matrix element of $H_{\rm spin}$ between states 
differing by a spin flip.  Thus the equivalent spin Hamiltonian is 
	\begin{eqnarray}
	\nonumber
	H_{\rm spin}&=&(-t)\sum_i \left(S_i^+ S^-_{i+1} + S^-_i S^+_{i+1}\right)\\
	&=&(-2t)\sum_i \left(S_i^x S^x_{i+1} + S^y_i S^y_{i+1}\right),
	\label{eq-Hspin}
	\end{eqnarray} 
which is the so-called spin-1/2 XX chain 
and is well-studied and exactly solvable.~\cite{Lieb}

\subsubsection{Mapping stripe states to spin states}

Lifted to act on Hilbert space, the map 
should take the mathematical form 
  \begin{equation}
     \SSS|n\rangle= |n_s\rangle, 
  \label{eq-SSSmap}
  \end{equation}
where $|n\rangle$ is a stripe state and $|n_s\rangle$ is a spin state. 

Correspondingly, the equivalent spin Hamiltonian should satisfy 
  \begin{equation}
	\SSS H| n\rangle =  H_{\rm spin} | \SSS n \rangle
  \label{eq-Hammap}
  \end{equation}
where $H$ is the spinless fermion or hardcore boson
Hamiltonian, Eq.~(\ref{eq-Ham}), and $|n\rangle$ is any single-stripe state.

However, in the fermion case these equations are inconsistent with an arbitrary 
site-ordering scheme, for Eq.~(\ref{eq-SSSmap}) did not take the
sign factors into account.  For many boundary conditions, there is
{\it no} way to define a unique function, on account of the large
loops in the state graph illustrated in Fig.~\ref{fig-sab}. 
That is, particles can be hopped successively (corresponding to 
a walk on the state graph) such that $|n\rangle$ returns to
the state $-|n\rangle$.  This gauge-like phase is familiar from
the spinor coherent states (where each spin direction is mapped
to a spin-1/2 ket), in which case a $2\pi$ rotation induces a 
$-1$ phase shift.  
 This is handled most transparently after a 
 further mapping of the spin chain to a fermion chain 
 (Sec.~\ref{sec-1Dfermion}, below). In that representation, 
 the phase factor can be associated with the hopping of 
 a one-dimensional fermion across the periodic boundary, 
 and it will be implemented by adding an artificial flux 
 that pierces the ring formed by the chain with 
 periodic boundary conditions (Sec.~\ref{sec-1stripephase}).

\subsubsection{Special case: zero gauge phase}

The many-to-one nature of the map (because $y\to y+2$ translation
of the stripe produces the same spin state) is easier to handle. 
Let us choose a rectangular system,
adopt the site-ordering conventions of Sec.~\ref{sec-basis}, 
and provisionally assume there is no gauge-like phase ambiguity.
We ask that if $T_{R_y}$ translates a state in the $y$ direction by $R_y$,
then $\SSS T_{R_y} |n\rangle=\SSS|n\rangle$.
One way to recover the notion of a unique map is to form 
Fourier states like Eq.~(\ref{eq-Bloch}), but only in the $y$ direction:
	\begin{equation}
	|n,k_y\rangle=\frac{1}{\sqrt{L_y}}\sum_{R_y}
	e^{-ik_y R_y}T_{R_y}|n\rangle.
	\label{eq-transstate}
	\end{equation}
The normalization factor here is always $1/\sqrt{L_y}$
because the $L_y$ translations in the sum cannot produce 
two identical states. From group theory we know
$\langle m, k_y'|H|n, k_y\rangle=0$ if $k_y\ne k_y'$, 
so states with different $k_y$ belong to disconnected blocks. 
In each block with a given $k_y$, there is a one-to-one
correspondence $|n,k_y\rangle \leftrightarrow \SSS|n\rangle$. 


Using Eq.~(\ref{eq-Hn}),
because $H$ commutes with translation $T_{R_y}$,
it is easy to show that
	\begin{equation}
	H|n,k_y\rangle=(-t)\sum_{m\in{\cal M}} s(n,m) |m,k_y\rangle.
	\end{equation}
For $q\in{\cal M}$, we then have 
	\begin{equation}
	\langle p,k_y|H|n,k_y\rangle=-s(n,p)t.
	\label{eq-Hpn}
	\end{equation}
But if we translate $p$ by $R_y$ so we have 
$T_{R_y}|p\rangle=\sigma |p'\rangle$, where $\sigma$ is
the fermion sign from translating
the state $p$ by $R_y$, then the translation
state $|p',k_y\rangle=\sigma e^{ik_y R_y}|p,k_y\rangle$.
Then we have,
	\begin{equation}
	\langle p',k_y|H|n,k_y\rangle=-s(n,p) \sigma e^{-ik_y R_y}t.
	\label{eq-Hp'n}
	\end{equation}

Eqs.~(\ref{eq-Hp'n}) and (\ref{eq-Hpn}) tell us that 
choosing $p'$ rather than $p$, the matrix element
$\langle p',k_y |H| n,k_y\rangle$
has an extra sign $\sigma$ and an extra phase
factor $e^{-ik_yR_y}$. However, in the spin-chain
language, we have,
$\langle \SSS p | H_{\rm spin} | \SSS n \rangle
=\langle \SSS p' | H_{\rm spin} | \SSS n \rangle$,
because $\SSS|p\rangle=\SSS|p'\rangle$. This means
that we should have $\sigma e^{-ik_y R_y}=1$ for all $R_y$.
Choose $L_y=3 \pmod 4$, 
then, as we have shown in Sec.~\ref{sec-1stripeBoseFermi}, 
every fermion hop 
in the one-stripe case produces a $+1$ sign,\cite{FN-signrule}
thus we have $s(n,p)=1$, and at the same time,
every $y$ translation gives $\sigma=1$.
In addition, if we choose $k_y=0$, then we have,
	\begin{equation}
	\langle p,0|H|n,0\rangle=-t,
	\label{eq-Hpn0}
	\end{equation}
for any $|p\rangle$ and $|n\rangle$ in their respective
translation classes. 

In summary, for $k_y=0$ and $L_y=3\pmod 4$,
we map the Fourier transformed single-stripe state
$|n,0\rangle$ to the spin-chain state $|\SSS n\rangle$,
then we have an identity of matrix elements 
 $\langle p, 0|H|n,0\rangle=\langle \SSS p|H_{\rm spin}|\SSS n\rangle$,
and thus Eq.~(\ref{eq-Hammap}) is satisfied.

For general $k_y\neq 0$ or $L_y \neq 3 \pmod 4$, this map
involves additional phase factors which will be
discussed in Sec.~\ref{sec-1stripephase} using
a further map to one-dimensional fermions
that we will describe in Sec.~\ref{sec-1Dfermion}.

We have checked that the 
eigenvalues and eigenvectors of the spin chain
system match those of the spinless fermion
and hardcore boson problems, using a spin diagonalization program
developed for another project.~\cite{Houle}


\subsection{Fermion representation of a stripe}
\label{sec-1Dfermion}

It is possible, in turn, to map each configuration of ``spins'' to
that of a one-dimensional (1D) lattice gas of particles:
every $\uparrow$ becomes an occupied site, every $\downarrow$ becomes a 
vacant one. 
The total number of ``up'' steps, i.e. fermions, is
$N_+\equiv (L_x+\uy)/2$  if the boundary vector along
the stripe is $(L_x,\uy)$. 
The $+y$ hop of a real particle on the square lattice 
(implying a $-y$ fluctuation of the stripe), 
translates to a spin exchange $\uparrow \downarrow \to \downarrow \uparrow$
in the XX chain, and finally to a +$x$ hop of a 1D particle. 

As is well-known,~\cite{LiebMattis}  in one dimension hardcore particles
may always be treated as fermions: if there is no path for them to
exchange, then the statistics has no physical meaning. 
One can describe the system as one-dimensional spinless fermions
which are {\it noninteracting}, as the Fermi statistics is already
sufficient to keep two particles from occupying the same site. 
The explicit relation of the stripe path $y(i)$  to the
1D spin and fermion representations is 
  \begin{equation}
  \label{eq-deltay}
        y(i+1)-y(i) \equiv 2S_{iz} \equiv 2 \hat{n}_i -1. 
  \end{equation}
where $\hat{n}_i \equiv c^\dagger_i c_i$. 
The effective one-dimensional Hamiltonian is 
   \begin{equation}
       {\cal H} = -t \sum _i (c_{i+1}^\dagger c_i + c_i^\dagger c_{i+1}) , 
   \end{equation}
thus the dispersion is 
   \begin{equation}
      -2 t \cos q, 
   \label{eq-1DEq}
   \end{equation}
where $q$ is the one-dimensional wavevector.  
We can construct exactly the ground state by filling
the lowest-energy plane-wave states, up to 
the one-dimensional Fermi vector $\nu\pi$ ($=\pi/2$ for
an untilted stripe); here
$\nu=N_+/L_x$ is the density of one-dimensional fermions, 
corresponding to a stripe slope $2 \nu -1$. 
All excited states correspond exactly to other ways of 
occupying the 1D fermion states. 

\subsubsection {Excitations of a stripe}
\label{sec-stripeexcite}

If a stripe is coarse-grained as a quantum-fluctuating string, then it can 
be approximated using a harmonic Hamiltonian, 
     \begin{eqnarray}
	\nonumber
     {\cal H} &=& {1\over 2} \int d x \left[ \rho \left(\frac {dy}{dt}\right)^2
                                    + K \left(\frac {dy}{dx}\right)^2\right]\\
          &=& \sum _q {1\over 2} \left[
              \frac{P_q^2}{\rhostar} + K q^2 {y_q^2}\right]. 
     \label{eq-string}
     \end{eqnarray}
Here $\rhostar$ is the effective mass density per unit length of the stripe, 
and the stiffness $K$ is analogous to a string tension. 
The long-wave excitations of such a ``string'' are the 
quantized capillary waves known as {\it ripplons}, which have dispersion 
     \begin{equation}
     \label{eq-ripplon}
         \omega=c |q|,  \hbox {\qquad with \qquad} c= \sqrt{K/\rhostar}. 
     \end{equation}
and $q$ is the one-dimensional wavevector. 

On the other hand,  in the microscopic representation
by 1D-fermions, the fundamental excitation is evidently a particle 
or hole, which corresponds to a mobile {\it kink} of the stripe.  
The ripplon is thus a composite excitation, a kink-antikink bound state.  
In the equivalent language of the spin-1/2 XX model, 
the ripplon maps to a magnon (hydrodynamic mode), 
while the kink maps to a spinon. 
The fractionalization of the spin-1 magnon into two
spin-1/2 spinons is a familiar fact, as this system is a 
special case of a Luttinger liquid.~\cite{Haldane}

We can use the Fermi sea representation to extract the 
parameters in Eq.~(\ref{eq-string}). 
As noted above, the 1D Fermi wavevector is $(1+dy/dx) \pi/2$ 
and by Eq.~(\ref{eq-1DEq}) 
the total energy is $\sumocc (-2t) \cos q$, 
where this notation means the sum over occupied
1D fermion states. 
Equating the energy density due to small tilts
to $\frac{1}{2} K (dy/dx)^2$, we obtain $K=\pi t/2$. 
Next, the ripplon velocity 
is the same velocity as the Fermi velocity,  $c=v_F=2t/\hbar$.  
(This is a standard fact of one-dimensional Fermi seas.)
With Eq.~(\ref{eq-ripplon}), that implies 
  \begin{equation}
  \label{eq-rhostar}
     \rhostar= \pi \hbar^2/ 8t.
  \end{equation}
 

\subsubsection{Stripe roughness}

The 1D Fermi sea representation also allows the exact computation
of stripe fluctuations as a function of $x$.  
It is straightforward to show 
(after rewriting in terms of wave operators) that 
at half filling, 
   \begin{equation}
   \label{eq-stripecorr}
       \langle (2\hat{n}_i-1)(2 \hat{n}_{i+r}-1) \rangle
            = \cases{1, &if $r=0$;\cr
             0, &     for even $r\neq 0$;\cr
        -\frac{4}{\pi^2 r^2}, & for odd $r$.\cr} 
   \end{equation}
We also know from Eq.~(\ref{eq-deltay}) that 
   \begin{equation}
       y(i+R)-y(i) = \sum _{j=i}^{R-1} (2 \hat{n}_j-1).
   \end{equation}
This and Eq.~(\ref{eq-stripecorr}) give that 
   \begin{eqnarray}
	\nonumber
       \langle [y(i+R)-y(i)]^2 \rangle  &=& 
        R+ \frac{8}{\pi^2} \sumodd
             \left( \frac{1}{r} - \frac{R}{r^2} \right)\\
       &\approx& {\rm const} + \frac{4}{\pi^2} \ln R
   \label{eq-stripefluc}
   \end{eqnarray}
Not surprisingly, the same result can be derived via
the zero-point intensities of the harmonic ripplon modes, 
of Eqs.~(\ref{eq-string}) and (\ref{eq-ripplon}). 

Eq.~(\ref{eq-stripefluc}) means that, due to the anticorrelations
evident in Eq.~(\ref{eq-stripecorr}), the transverse deviations of
the stripe grow slowly with length.  Thus, if it moves 
transversely over larger distances, 
the stripe can often be approximated as a single rigid object.  
This approximation will be invoked in several
sections to explain how energy splittings depend on the 
system size $L_y$ in the transverse direction. 

\subsection {Phase factors and stripe effective mass}
\label{sec-1stripephase}

The mappings described in Sec.~\ref{sec-1Dfermion}
are less trivial than they appeared, due to two related facts. 
The first fact is that the mapping of stripe to spins (or to 1D particles) 
is many-to-one, as already noted in Sec.~\ref{sec-maptospin}. 

The second fact is that the statistics of 1D hardcore particles 
is not quite as irrelevant as suggested in Sec.~\ref{sec-1Dfermion}:
they {\it can} be permuted
by moving them in the $x$ direction through the periodic boundary conditions.
As is well known, this induces additional phase factors in
a finite system. 
Consider, for example, a sequence of $-y$ hops along a stripe, such that
exactly one particle hops in each column. The net effect on the real particle
configuration is to translate the stripe by (0,2); in the fermion case, 
a sign factor $\sigma_{\rm col}^{h}$ is also picked up, where $h$ is the 
number of columns in which the real particle hop crossed the cell boundary, and 
Eq.~(\ref{eq-sigmacol})
is the fermion sign picked up from
the resulting rearrangement of creation operators among sites in a column. 
(Note that, here and for the rest of this discussion, we limit consideration
to lattices in which the second boundary vector is $(0, L_y)$.)
Meanwhile, the same sequence of real particle hops
maps to a  cyclic permutation of the 1D particles in the $+x$ direction --
exactly one $+x$ hop occurs on every bond along the chain. 

If the wavevector of our eigenstate is taken as $(k_x,k_y)$, the stripe configuration
shifted by $(0,2)$ ought to have an amplitude 
in the wavefunction $e^{i2 k_y}$ times the amplitude of the original configuration.  
In a 1D fermion system, however, the actual phase factor is 
$(-1)^{N_+-1}$, 
i.e. the number of fermions
that are cyclically permuted. 
We can account for all the phase factors by 
modifying the one-dimensional chain so that a particle hopping around its
periodic boundary condition picks up a phase $e^{i\phi}$
in other words by inserting a flux $\phi$ into the ring,~\cite{FN-1Dring}
where 
        $\phi= 2 k_y + (N_+-1)\pi$
is sufficient to account for the two phases mentioned above, 
in the boson case; in the fermion case, the
an additional term $\frac{1}{2}L_x(L_y+1)\pi$ is needed.~\cite{FN-Lygauge}
Finally it is equivalent, modulo $2\pi$, 
to replace $L_x$ by $\uy$ in the last term:
   \begin{equation}
        \phi= 2 k_y + (N_+-1)\pi + [\frac{1}{2}\uy(L_y+1)\pi]_{\rm fermions},
   \label{eq-phi1stripe}
   \end{equation}
where the last term is included only in the fermion case, and
even then it is zero in a rectangular system. 
[Notice that $L_y$ is always odd when the boundary vector 
is $(0,L_y)$ transverse to the stripe, and the other
boundary vector $(L_x,\uy)$ must be even.] 
It follows that the allowed 1D wavevectors are $(2 \pi m+ \phi)/L_x$, 
where $m$ is any integer.  

We get $\phi=0$ when 
the number $N_+$ of 1D fermions is odd and $\phi=\pi$ when $N_+$
is even; either way, 
the 1D fermions occupied in the ground state
are placed symmetrically about $q=0$. 
In any eigenstate, the real wavevector component is 
$k_x = \sumocc q$, 
thus $k_x=0$ for the stripe ground state
in any rectangular system. 
It is not hard to compute $\sumocc (-2t) \cos q$ 
to obtain the (exact) total ground state energy 
   \begin{equation}
   \label{eq-stripeE0}
      E_{\rm stripe}(0,0) = -2 t/\sin (\pi/L_x), 
   \end{equation}
which implies the stripe energy is $-2t/pi$ per unit length 
in the thermodynamic limit.~\cite{Mila}
All our numerical results for stripe states agree
with Eq.~(\ref{eq-stripeE0}) (see Table~\ref{t-excite}). 

When, as mentioned in Sec.~\ref{sec-stripeexcite}, 
we model the stripe direction as a free particle moving
in one dimension, it is useful to know its effective mass $m^*$.  
The minimum-energy state of a single stripe
with $k_y \neq 0$ consists of rigid 
motion in that direction, so for small $k_y$ we expect
   \begin{equation}
   \label{eq-mstripe}
      E_{\rm stripe}(0,k_y) - E_{\rm stripe}(0,0) \cong
      \frac {\hbar^2 {k_y}^2} {2 m^*} .
   \end{equation}
In the 1D fermion representation, 
this state is constructed by shifting
the ground state Fermi sea by
a wavevector $2 k_y/L_x$, 
as follows from the first term in the 
phase shift Eq.~(\ref{eq-phi1stripe}). From this one obtains
$E_{\rm stripe}(0,k_y) =$ 
$E_{\rm stripe}(0,0)\cos(2 k_y/L_x).$
Combining that with Eqs.~(\ref{eq-stripeE0}) and (\ref{eq-mstripe}), 
one obtains 
   \begin{equation}
   \label{eq-mstripeLx}
         m^* = \frac{\hbar^2 L_x^2}{8t} \sin (\pi/L_x).
   \end{equation}
For a long stripe, Eq.~(\ref{eq-mstripeLx})
implies an effective mass density 
$\rhostar \equiv \lim_{L \to \infty} m^* /L_x $ 
in agreement with Eq.~(\ref{eq-rhostar}). 

To obtain the effective mass $m^*$ from
exact diagonalization data, we 
use the smallest $k_y=2\pi/L_y$ with a large
$L_y=101$. Eq.~(\ref{eq-mstripe}) gives the following expression
to extract $m^*$ numerically, 
	\begin{equation}
	m^*\cong \frac{(2\pi/L_y)^2}{2(E_{stripe}(0,2\pi/L_y)-E_{stripe}(0,0))},
	\label{eq-mexcite}
	\end{equation}
The results are shown in Table~\ref{t-excite};
they agree perfectly with 
Eq.~(\ref{eq-mstripeLx}). 

\begin{table}[ht]
        \centering
        \caption{
Stripe effective mass $m^*$ [Eq.~(\ref{eq-mexcite})]
using single-stripe excited state energies with $\kk=(0,k_y)$, 
with $L_y=101$. 
}
        \begin{tabular}{cccc} 
$L_x$ & $E_{stripe}(0,0)$ & $E_{stripe}(0,2\pi/101)$ & $m^*$ \\\hline
4  & -2.8284271 & -2.8270590 & 1.4143276 \\
6  & -4.0000000 & -3.9991400 & 2.2500806 \\
8  & -5.2262519 & -5.2256198 & 3.0615292 \\
10 & -6.4721360 & -6.4716350 & 3.8627623 \\
12 & -7.7274066 & -7.7269913 & 4.6587845 \\
14 & -8.9879184 & -8.9875635 & 5.4517988 \\
        \end{tabular}
        \label{t-excite}
\end{table}

It should be noted that the above calculation of the
phases is entirely equivalent to that of 
Green and Chamon~\cite{green00,green02}
(where it is applied to a stripe on
a triangular lattice), except that we include the
case that the particles are fermions.

\section{One Hole on a Stripe}
\label{sec-onehole}

\begin{figure}[ht]
	\centering
	\includegraphics[width=\linewidth]{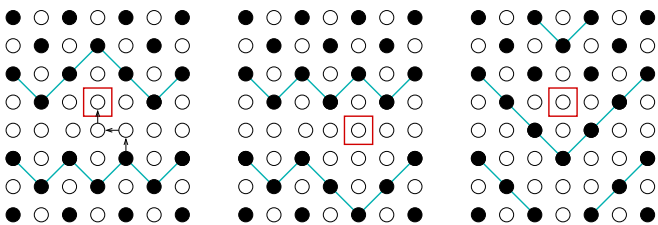}
	\caption{
	A $6\times 7$ system with a stripe and a hole.
	The hole is indicated in the box.
	It is initially lying on the stripe (leftmost state).
	The arrows indicate the particles hops that result
	in the hole moving along the stripe to the right (middle state).
	It is also possible for the stripe to fluctuate
	and leave the hole immobile (rightmost state).
	}
	\label{fig-holeonstripe}
	\end{figure}

For a system with a single stripe, we have seen
that the particle motion is strictly limited to sliding
in the direction perpendicular to the stripe,
and this enables us to map our two-dimensional system
to a one-dimensional spin chain and 
solve it exactly. With holes added to the stripe state,
a number of new motions are allowed. 
In Fig.~\ref{fig-holeonstripe} we show a $6\times 7$
system with a stripe and a hole.
We see that the hole can now move along the stripe
and also, the stripe can fluctuate and leave the
hole stranded (i.e., immobile). In this section we study
the one-hole-with-a-stripe problem. 

Motion of holes has been analyzed on domain walls in the anisotropic $t$-$J$
model (the $t$-$J_z$ model with some added terms), using essentially 
analytical techniques.~\cite{Tch}
The most interesting observation in Ref.~\onlinecite{Tch}
is that the charge-carrying ``holon''
is associated with a kink of the wall, and 
thus carries a transverse ``flavor'' (reminiscent of
our model's behavior in Sec.~\ref{sec-holekink}). 
A hole on our stripe differs in an important way from
the holes on a stripe in the $t$-$J_z$ model.~\cite{Tch,CherNeto00}
First, in our case the reference stripe already has 1/2 hole
per unit length, whereas in Refs.~\onlinecite{Tch} and
\onlinecite{CherNeto00} it is a plain domain wall without holes. 
Second, $J_z/t$ was not too large in those models, so that 
hopping of the hole several steps away from the stripe has
a noticeable contribution in the wavefunction, while it is
unimportant in our case.

>From diagonalization, we observe that for
systems with rectangular boundaries, the 
boson and fermion one-hole-with-a-stripe spectra
are identical. The numerical method 
described in Sec.~\ref{sec-statistics}
is used to check whether the function
$\sigma(\alpha)$ is well defined using
the state graph Fig.~\ref{fig-tree}.
For a $L_x\times L_y$ system with $L_x$ even
and $L_y$ odd, the one-hole-with-a-stripe
state has $M=(L_x-1)L_y/2-1$ particles.
For the one-hole systems we checked, for example, $4\times 9$
with $M=15$, $4\times 11$ with $M=19$, 
and $6\times 7$ with $M=17$,
$\sigma(\alpha)$ is always well defined.~\cite{Henley-stripehole}

\subsection{Energy dependence on $L_y$}
\label{sec-1holeLy}

When a hole is added to a stripe, more hops are allowed
and the state gains kinetic energy, so the energy
is lower than that of the single stripe of
the same length. We define the energy difference,
	\begin{equation}
\Delta(L_x,L_y)\equiv E_{hole}(L_x,L_y)-E_{stripe}(L_x),
	\label{eq-Delta}
	\end{equation}
where $E_{hole}(L_x,L_y)$ is the ground state energy of 
one hole with a stripe on a $L_x\times L_y$ 
lattice and $E_{stripe}(L_x)$ the ground state energy of a single
stripe with length $L_x$. Here $E_{hole}(L_x,L_y)$ is the same for
bosons and fermions, as we discussed above, and
$E_{stripe}(L_x)$ does not depend on $L_y$ (see Eq.~(\ref{eq-stripeE0})).

A plot of $\Delta(L_x,L_y)$ 
vs $L_y$ shows a fast decay in $L_y$ so we try
the following exponential fitting function,
	\begin{equation}
	\Delta(L_x,L_y)={\tilde \Delta}(L_x)-A(L_x)e^{-L_y/l(L_x)},
	\label{eq-Dxy}
	\end{equation}
where ${\tilde \Delta}(L_x)=\Delta(L_x,\infty)$, 
$A(L_x)$, and $l(L_x)$ are fitting
parameters that depend on the length of the stripe 
$L_x$. (We will investigate the dependence
on $L_x$ of ${\tilde \Delta}(L_x)$ later in Sec.~\ref{sec-oneholeLx}.)
We choose a minus sign in front of $A(L_x)$ because
$\Delta(L_x,L_y)<{\tilde \Delta}(L_x)$; 
in Eq.~(\ref{eq-Dxy}), $A(L_x)$ is positive. 

\begin{figure}[ht]
	\centering
	\includegraphics[width=\linewidth]{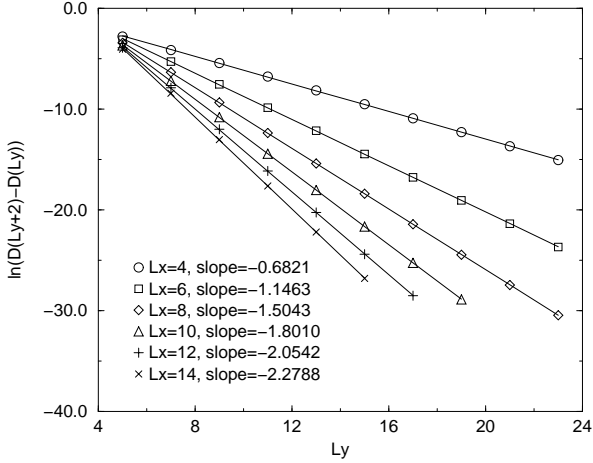}
	\caption{
	$\ln(\Delta(L_x,L_y+2)-\Delta(L_x,L_y))$ vs $L_y$
	for $L_x=4,6,8,10,12,14$, where $\Delta(L_x,L_y)=E_{hole}-E_{stripe}$,
	Eq.~(\ref{eq-Delta}).
	The slope for each $L_x$ curve is $-1/l(L_x)$ in Eq.~(\ref{eq-expfit}).
	}
	\label{fig-Ly_hole_ln}
	\end{figure}

This fitting form Eq.~(\ref{eq-Dxy}), 
suggests the following linear regression check,
	\begin{equation}
	\ln(\Delta(L_x,L_y+2)-\Delta(L_x,L_y))
	=C-\frac{L_y}{l(L_x)},
	\label{eq-expfit}
	\end{equation}
where $C$ is a constant that depends on $A$, $l$, and $L_x$,
but not on $L_y$. In Fig.~\ref{fig-Ly_hole_ln} we plot 
$\ln(\Delta(L_x,L_y+2)-\Delta(L_x,L_y))$
vs $L_y$ for $L_x=4,6,8,10,12,14$ and $L_y=5,7,9,...$.
The linear fit is excellent for all data sets.

\subsection{Stripe potential well and effective mass}
\label{sec-doublewell}

How can we account for the exponential fitting
form, Eq.~(\ref{eq-Dxy})? 
Let us consider the hole fixed 
at some position and the stripe meandering in
$y$ direction. In Fig.~\ref{fig-holeonstripe}, we have
shown that the hole can be in contact with the stripe
or the stripe can fluctuate away, leaving the
hole behind and immobile. When the stripe is in contact
with the hole, the energy is lower than the energy of 
a single stripe $E_{stripe}$, which is also the energy when 
the stripe is separated from the hole. 
Because we have periodic boundary conditions
in the $y$ direction, we can use a periodic array of potential wells
to model the $y$ motion of the stripe; the well is at the 
position of the hole. 

	\begin{figure}[ht]
	\centering
	\includegraphics[width=0.7\linewidth]{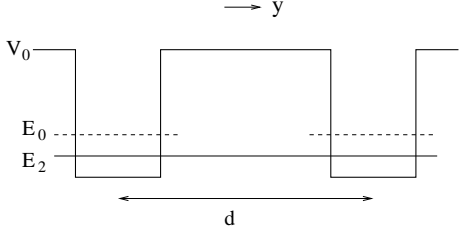}
	\caption{
	Potential-well and barrier potential used to understand the
	exponential decay of one-hole energy in $L_y$ in Eq.~(\ref{eq-Dxy}).
	$y$ is the stripe position in the direction
	perpendicular to the stripe, $V_0$ is the depth of the well,
	$E_0$ the ground state energy
	of one isolated well, $E_2<E_0$ the ground state energy
	including the tunneling between wells (symmetric state), and 
	$d$ the separation between two wells.
	$E_2-E_0$ decays exponentially in $d$ as in 
	Eq.~(\ref{eq-2wells}).
	}
	\label{fig-well}
	\end{figure}

In Fig.~\ref{fig-well} we show potential wells separated by
a barrier of thickness $d$. $E_0$ is the ground state energy of an isolated well,
and $E_2$ the ground state energy of the multiple-well system,
which corresponds to a symmetric state and
is lower than $E_0$. From standard quantum mechanics
textbooks (see e.g., Ref.~\onlinecite{Park}), we know the 
amplitude to tunnel between adjacent wells 
difference decays exponentially with their separation,
	\begin{equation}
	E_2-E_0=-A e^{-d/l}, \quad\mbox{where,}\quad
\frac{1}{l}=\frac{\sqrt{2m(V_0-E_0)}}{\hbar},
	\label{eq-2wells}
	\end{equation}
where $m$ is the mass of the particle in the well, 
and the constant $A$ is positive (because the 
wavefunction of $E_0$ is symmetric in one well).

As far as our hole-with-a-stripe problem is concerned,
the well separation $d$ is $L_y$,
the barrier height $V_0$ is $E_{stripe}$, the
ground state energy of one isolated well $E_0$ is
$E_{hole}(L_x,L_y=\infty)$, and the
ground state $E_2$ of the system with tunneling
is $E_{hole}(L_x,L_y)$. Therefore, Eq.~(\ref{eq-2wells})
for the wells with tunneling translates into the following
equation for our hole-with-a-stripe problem,
	\begin{equation}
	E_{hole}(L_x,L_y)-E_{hole}(L_x,\infty)
	=-A(L_x) e^{-L_y/l(L_x)}.
	\label{eq-exp}
	\end{equation}
Using the definition for $\Delta(L_x,L_y)$, Eq.~(\ref{eq-Delta}), 
we see that Eq.~(\ref{eq-exp}) is exactly the fitting form we used
before, Eq.~(\ref{eq-Dxy}). In addition, the inter-well tunneling
amplitude Eq.~(\ref{eq-2wells}) gives us a way to calculate
the effective mass $m^*(L_x)$ of the stripe of length $L_x$,
	\begin{equation}
	\frac{1}{l(L_x)}=
\sqrt{2m^*(L_x)(E_{stripe}(L_x)-E_{hole}(L_x,\infty))},
	\end{equation}
where we have set $\hbar=1$. We get
	\begin{equation}
	m^*(L_x)=\frac{1/l^2(L_x)}{2(E_{stripe}(L_x)-E_{hole}(L_x,\infty))}.
	\label{eq-m}
	\end{equation}
Using the linear fitting slopes in Fig.~\ref{fig-Ly_hole_ln}
(that are $-1/l(L_x)$), we can compute $m^*(L_x)$ using
Eq.~(\ref{eq-m}), and our results are shown in Table~\ref{t-m}.
The effective mass results are consistent with that obtained
from the single-stripe energy dispersion relation
in Table~\ref{t-excite}.

\begin{table}[ht]
        \centering
        \caption{
Stripe effective mass $m^*$ calculated from Eq.~(\ref{eq-m})
using the potential-well model for
the one-hole-with-a-stripe problem.
$l(L_x)$ is the decay
length in Eq.~(\ref{eq-Dxy}) obtained from linear fitting in 
Fig.~\ref{fig-Ly_hole_ln};
$E_{hole}(L_x,L_y)$ is used to approxmate
$E_{hole}(L_x,\infty)$ using the large $L_y$ listed
in the table (as the exponential decay of $E_{hole}(L_x,L_y)$ 
in $L_y$ is fast, see Sec.~\ref{sec-1holeLy});
and $m^*(L_x)$ is the effective
mass of the length-$L_x$ stripe calculated using
Eq.~(\ref{eq-m}).
$E_{stripe}(L_x)$ has appeared in Table~\ref{t-excite}.
}
        \begin{tabular}{ccccc} 
$L_x$ & $L_y$ & $E_{hole}(L_x,L_y)$ & $l(L_x)$ & $m^*(L_x)$ \\\hline
4 & 25  & -3.00000009 & 1.4659 & 1.3561 \\
6 & 25  & -4.29850460 & 0.8724 & 2.2009 \\
8 & 25  & -5.60276427 & 0.6648 & 3.0050 \\
10 & 21 & -6.89914985 & 0.5553 & 3.7979 \\
12 & 19 & -8.18958713 & 0.4868 & 4.5652 \\
14 & 17 & -9.47602246 & 0.4388 & 5.3197
        \end{tabular}
        \label{t-m}
\end{table}

The main physical significance of the result 
in this section is its
implications for the anistropic transport of holes in a 
stripe array.~\cite{anisotransport,ZaanenScience}
The conductivity transverse to the stripes depends completely
on mechanisms which transfer holes from one stripe
to the next. 
The high energy of the ``stranded'' state, in which the hole is
immobile and off the stripe, 
means that holes are mainly transferred when the stripes
collide, either directly at contact or else by a delayed transfer, 
such that the hole spends a short time in a stranded state
until the second stripe fluctuates to absorb it.

\subsection{Hole dispersion and $L_x$ dependence}
\label{sec-oneholeLx}

\begin{figure}[ht]
	\centering
	\includegraphics[width=\linewidth]{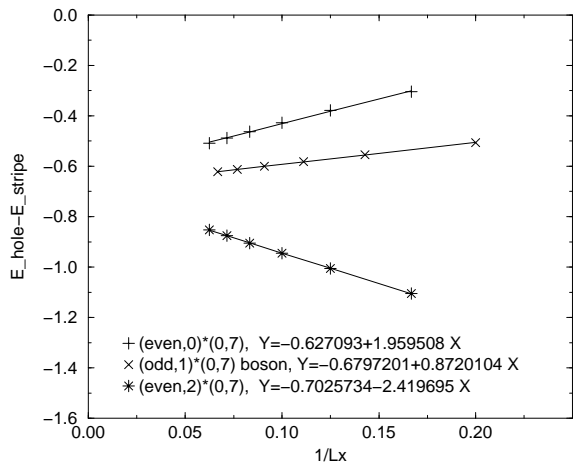}
	\caption{
	$\Delta(L_x,L_y)=E_{hole}-E_{stripe}$ 
	as a function of $1/L_x$ for three classes
	of lattices $(L_x,b)\times(0,7)$ with $b=0,1,2$.
	}
	\label{fig-onehole_inverse}
	\end{figure}

In this section we fix $L_y$ and study the
dependence of $\Delta(L_x,L_y)$ on
the length of the stripe $L_x$. 
In Fig.~\ref{fig-onehole_inverse} we plot the
energy difference $\Delta(L_x,L_y)$
for $(L_x,b)\times(0,7)$
systems with $b=0,1,2$.\cite{FN-bosonLx}

We see from Fig.~\ref{fig-onehole_inverse} the following
fitting function (justified in Sec.~\ref{sec-holekink}) works well for both
rectangular and tilted lattices,
	\begin{equation}
	{\tilde \Delta}(L_x)=\Delta+\frac{C(b)}{L_x}.
	\label{eq-inversefit}
	\end{equation}
This fitting form enables us to extrapolate the energy gap
formed by adding one hole to an infinitely long ($L_x$)
stripe. The intercepts of the three
curves for the three different classes of lattices
all approach $\Delta=-0.66$. Later, 
in Sec.~\ref{sec-stability},
we will study the stability of an array of stripes
and we will use the value for $\Delta$ because 
we want to know whether holes added to a stripe
stick to the stripe to form a wide stripe or a new stripe.
$\Delta$ is energy lowered by adding a hole to
a stripe and will be relevant there. It is a binding
energy in the sense that one hole off the stripe (immobile) has
zero energy.

A hole is mobile along a stripe; 
one expects its dispersion relation to be 
   \begin{equation}
       E(q) = \Delta + \frac{\hbar^2}{2 m^*_h} q^2
   \label{eq-Eqstripehole}
   \end{equation}
on a long stripe oriented in the $x$ direction, 
where $q$ is the hole's wavevector. 
We can estimate the effective mass in Eq.~(\ref{eq-Eqstripehole}) from
the numerics, if the lowest-energy state 
with system wavevector ${\bf k}= (q,0)$ is produced by 
boosting the hole to wavevector $q$. 
At small $q$, this has a much smaller excitation energy than 
the ripplons (stripe excitations) which have linear dispersion 
[Eq.~(\ref{eq-ripplon})].  Indeed, the numerical spectra
for systems with $L_x=10$ and $L_y=5$ show these contrasting
dispersions for a stripe with and without a hole. 
Fitting the difference $E(q)-E(0)$ to $\hbar^2/ m^*_h (1-\cos q)$, 
as a plausible guess, 
we estimate $\hbar^2/2 m^*_h \approx 0.3t$; the effective
mass $m^*_h$ of a hole bound to a stripe is thus about six times 
as big as that of a free particle in an empty background
(for which $\hbar^2/2 m^* = 2t$). 

\subsection {Hole and stripe steps}
\label{sec-holekink}

How does the hole interact with the stripe fluctuations?
In some models, hole hopping is suppressed if the stripe
is tilted away from its favored direction. In that case, 
the ``garden hose effect'' is realized:~\cite{Na97}
stripe fluctuations are suppressed when the hole is
present, so as to optimize the kinetic energy of the hole
on the stripe.~\cite{FN-Nagaoka}  

\begin{figure}[ht]
	\includegraphics[width=0.3\linewidth]{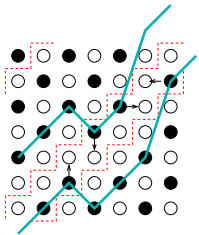}
  \caption{
A $7\times 7$ system is shown, at one particle less 
than maximum filling.  All motions in this system 
occur within the strip bounded by dashed lines, which is equivalent
to a $14\times 1$ chain of spinless $V=\infty$ fermions.  
The stripe edge is indicated by thick solid lines.} 
\label{fig-45hole}
\end{figure}

However, in our model, a stripe tilt actually enhances hole 
hopping.  This is clear in the extreme case of a stripe at 
slope $+1$, as occurs at maximum filling in the $L\times L$ square
system with $L$ odd: the bare stripe has zero energy. 
When the system is doped by placing one hole along an edge of the stripe
(see Fig.~\ref{fig-45hole}), 
the accessible configurations are equivalent to a one-dimensional 
chain of the same model, with length $2L$ and $L-1$ particles on it. 
The one-dimensional chain has two domain walls, one of which 
appears as a down-step of the stripe, the other of which appears
as an up-step bound to a hole on the stripe. 
In the limit of large $L$, each domain wall has kinetic energy $-2t$.
The total energy is $-4t$, or about seven times larger than the
energy difference $-\Delta$ for a stripe with no tilt. 

This picture is supported by the trend of the stripe energy
with one hole, in the presence of a boundary
vector $(L_x,b)$ that forces a tilt. For $L_x=5$ to $8$, 
the minimum energy, as a function of $b$,  occurs around $2.5$.
[The measured minimum is at $b=2$ in the case that $L_x$ and $b$ are even, 
at $b=3$ in the odd case, while the second-lowest energies are
at $b=4$ and $b=1$, respectively.]
This is consistent with the idea that the hole on a stripe 
binds with some up steps~\cite{FN-holestep}
which total $b_0$, where $b_0 \approx 2.5$.

This would imply that holes tend to come in ``up'' and ``down'' flavors;
this is not an actual quantum number, since
the ``steps'' considered here (in contrast to ``kinks'' mentioned in
Sec.~\ref{sec-stripeexcite}) are not quantized objects. 
Our hole-step binding has a similar origin as that
in Ref.~\onlinecite{Tch}: the local configuration is the same
as a fragment of stripe rotated $90^\circ$ from the main stripe. 

A corollary of the hole-step binding is that
a hole should force an extra tilt $|b-b_0|/L_x$
on the remaining portion of the stripe. Thus the tilt energy
in Eq.~(\ref{eq-string}) contributes a size dependence $C(b)/L_x$, 
where 
   \begin{equation}
           C(b) = \frac {1}{2} K [(b-b_0)^2 -b^2] = K b_0 (b_0/2-b),
   \label{eq-holeLx}
   \end{equation}
that is 4.9, 1.0, and -2.9 for $b=0$, 1, and 2 respectively. 
This is consistent with the clear $1/L_x$ size dependence
we found numerically (see Fig.~\ref{fig-onehole_inverse});
the fitted coefficients for $b=1$ and $2$ are
in reasonable agreement with $C(b)$ as predicted by Eq.~(\ref{eq-holeLx}).
(Presumably $C(0)$ is reduced from our prediction
because the hole is not always, or not usually,
bound to an up-step in the case $b=0$.)

\section{Two Holes on a Stripe}
\label{sec-twoholes}

When two holes meet on a stripe, they create enough room 
that particles 
can exchange with each other, so the boson
and fermion energy spectra are no longer the same.
As in the one-hole case, we study the energy of 
the two-holes-with-a-stripe problem as a function
of the two directions of the lattice.

\subsection{Energy dependence on $L_y$}
\label{sec-2holeLy}

As in Eq.~(\ref{eq-Delta}) for the one-hole case,
we define for the $L_x\times L_y$ lattice
the energy difference between the case of
two holes with a stripe and that of a single stripe,
	\begin{equation}
	\Delta_2(L_x,L_y)=E_{2holes}(L_x,L_y)-E_{stripe}(L_x),
	\label{eq-Delta2}
	\end{equation}
where the subscript 2 denotes the two-hole case. (Strictly speaking,
we should write $\Delta_2^{b,f}(L_x,L_y)$ and $E_{2holes}^{b,f}(L_x,L_y)$
because these quantities are not the same for boson and fermion cases.
Here, without the superscripts, they stand for both cases.)
Plots of $\Delta_2(L_x,L_y)$ vs $L_y$ for bosons and fermions 
show fast decay similar to that in the 
one-hole problem in Sec.~\ref{sec-onehole},
except here the boson and fermion curves
approach different values at large $L_y$.

As in Eq.~(\ref{eq-Dxy}) for the one-hole problem, we write
	\begin{equation}
	\Delta_2(L_x,L_y)={\tilde \Delta}_2(L_x)-A_2(L_x)e^{-L_y/l_2(L_x)},
	\label{eq-Dxy2}
	\end{equation}
$\ln(\Delta_2(L_x,L_y+2)-\Delta_2(L_x,L_y))$ 
vs $L_y$ for bosons and fermions have been plotted
with $L_x=4,6,8,10$.
The exponential dependence is checked nicely as in
Fig.~\ref{fig-Ly_hole_ln} for the one-hole case,
and the slopes are obtained and will be used 
for calculating the effective mass in the next section.

\subsection{Stripe potential well and effective mass}
\label{sec-doublewell2}

The resemblance of our treatment of the two-holes-with-a-stripe
problem in Sec.~\ref{sec-2holeLy} with that of the one-hole
problem in Sec.~\ref{sec-1holeLy} prompts us to ask whether
the two-hole problem can be studied using a one-dimensional
effective potential like the potential wells used in 
Sec.~\ref{sec-doublewell}. Here with two holes, we have a more
complicated problem because the relative positions of 
the two holes and the stripe can have three cases:
two holes on the stripe (with energy $E_{2holes}(L_x,\infty)$), 
one hole on the stripe with one 
isolated hole (with energy $E_{1hole}(L_x,\infty)$), 
or one stripe with two isolated holes (with energy $E_{stripe}(L_x)$). 

As an approximation, we consider the deepest well ($E_{2holes}(L_x,\infty)$)
only because that is the most probable position to find the particle. Then the 
effective mass equation Eq.~(\ref{eq-m}) is modified
to become,
	\begin{equation}
	m^*_2(L_x)=\frac{1/l^2_2(L_x)}{2(E_{stripe}(L_x)-E_{2holes}(L_x,\infty))},
	\label{eq-m2}
	\end{equation}
where $l_2(L_x)$ comes from the linear fitting slopes. 
The results for $m^*_2$
from this two-hole-with-a-stripe calculation are in Table~\ref{t-m2}.
They are comparable to the one-hole results in Table~\ref{t-m}.

\begin{table}[ht]
        \centering
        \caption{
Stripe effective mass $m^*_2$ calculated from Eq.~(\ref{eq-m2})
using the double-well potential model for
the two-holes-with-a-stripe problem.
$l_2(L_x)$ is the decay
length in Eq.~(\ref{eq-Dxy2}) obtained from 
linear fitting.
$E_{2holes}(L_x,L_y)$ is used to approxmate
$E_{2holes}(L_x,\infty)$ using the large $L_y$ listed
in the table; and $m^*_2(L_x)$ is the effective
mass of the length-$L_x$ stripe calculated using
Eq.~(\ref{eq-m2}). The superscripts $b$ and $f$ denote
bosons and fermions respectively. $E_{stripe}(L_x)$
has appeared in Table~\ref{t-excite}.
}
        \begin{tabular}{cccccccc} 
$L_x$ & $L_y$ & $E_{2holes}^b$ & $E_{2holes}^f$ 
& $l_2^b$ & $l_2^f$ & ${m_2^*}^b$ &  ${m_2^*}^f$ \\\hline
4  & 19 & -3.896952 & -3.818556 & 0.5919 & 0.6135 & 1.3358 & 1.3417 \\
6  & 19 & -5.238613 & -5.204899 & 0.4317 & 0.4386 & 2.1664 & 2.1574 \\
8  & 19 & -6.524636 & -6.510024 & 0.3621 & 0.3643 & 2.9375 & 2.9349 \\
10 & 15 & -7.797992 & -7.790359 & 0.3167 & 0.3274 & 3.7605 & 3.5377 \\       
\end{tabular}
        \label{t-m2}
\end{table}

\subsection{Energy dependence on $L_x$
and stripe-step binding}
\label{sec-stripe90}

\begin{figure}[ht]
	\centering
	\includegraphics[width=\linewidth]{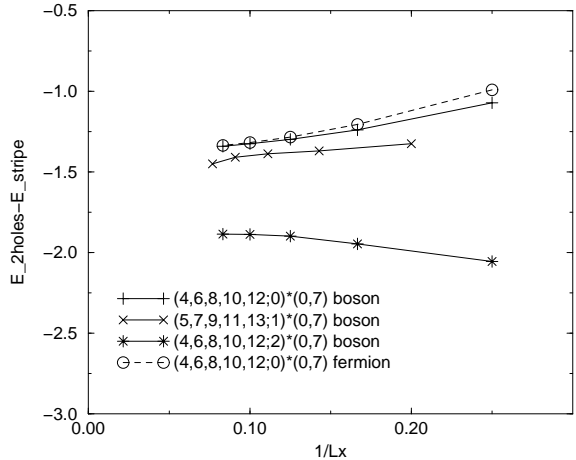}
	\caption{
	$\Delta_2(L_x,L_y)=E_{2holes}-E_{stripe}$ 
	as a function of $1/L_x$ for three classes
	of lattices $(L_x,b)\times(0,7)$ with $b=0,1,2$.
	}
	\label{fig-twoholes_inverse}
	\end{figure}

To compare with Fig.~\ref{fig-onehole_inverse}
for one hole with a stripe, in Fig.~\ref{fig-twoholes_inverse}
we plot $\Delta_2(L_x,L_y)$, for three classes of lattices
$(L_x,b)\times(0,L_y=7)$ with $b=0,1,2$, vs $1/L_x$.
Here the $1/L_x$ fit, as in Eq.~(\ref{eq-inversefit})
for the one-hole problem, is no longer good.
But we can still extrapolate the energy gap of the
two-hole state and the stripe state for an infinitely
long stripe. For the $(L_x,0)\times(0,7)$ 
lattices, the gap is extrapolated to be $-1.4$ approximately, 
for bosons and fermions.
Comparing to the single hole gap of $-0.65$, 
we see that binding between holes are not strong.
This information will be useful when we consider 
stripe-array formation.

The size dependence of two holes on one stripe
can be interpreted by the notion (Sec.~\ref{sec-holekink})
that holes are bound
to up- or down-steps. 
When the boundary vector is $(L_x,b)$ with $b=0$ or $1$, the stiffness cost will be 
minimized if the holes adopt canceling flavors ``up'' and
``down'' (so long as $b< 2b_0 \approx 5$). 
This may explain why the $b=0$ curves in 
Fig.~\ref{fig-twoholes_inverse} seem to have a 
weak coefficient of $1/L_x$. 
The fact that $\Delta_2 \approx 2 \Delta_1$ indicates 
weak interaction between the holes, and that can only
happen if the holes (of opposite flavors) tend to repel. 

However, on reflection it will be noticed 
that, in a sufficiently large system,
added holes can be ``condensed'' to create segments of stripe 
oriented in the $\pm \hat{y}$ direction. 
These ``$90^\circ$'' segments will surely appear in the ground
state of one stripe in a large system, as the energy per hole to
increase the net stripe length is far lower than the energy 
to add a free hole hopping along the stripe. 
Thus, holes do attract,  on a sufficiently long stripe.  
That does not contradict our numerical observation of
repulsion: as just explained, the constraint of 
zero net kink ``flavor'' on a stripe forces two holes to
adopt opposite flavors, which repel, 
whereas the holes that form a single $90^\circ$ segment
all have the same flavor.

Two effects compete with the formation of $90^\circ$ segments 
in the case of short stripes or
small numbers of added holes:
(i) Say the stripe contains {\it one} $90^\circ$ segment with $h$ holes;
that extends $\pm 2h$ ({\it not} $\pm b_0h$) in the $y$ direction, 
which forces the remainder of the stripe to have a slope $\mp 2h/L_x$ and the
associated tilt energy Eq.~(\ref{eq-string}). 
(ii) Say the stripe contains {\it two} $90^\circ$ segments, 
with the opposite directions: no net slope is forced, 
but twice as many $90^\circ$ corners are
present, and presumably each $90^\circ$ bend costs energy
(since it suppresses stripe fluctuations). 

The implication is that the $b=0$ and $b=1$ curves in
Fig.~\ref{fig-twoholes_inverse} are heading towards a well-defined asymptote
at $-2\Delta_1$, corresponding to a pair of repelling, 
opposite-flavor holes.  However, at sufficiently
long $L_x$, they must cross over to a different curve
with a lower asymptote, 
corresponding to a pair of attracting, same-flavor holes, 
bound into an incipient $90^\circ$ segment. 
We conjecture that the entire $b=2$ curve is in the latter regime, 
while the downturn of the $b=1$ curve at the largest $L_x$ suggests
it is beginning to make the crossover. 


\section{Two and More Stripes}
\label{sec-twoandmore}

Starting from this section, we study the interaction among stripes. 
We use the same diagonalization
program for the problem of one stripe and one stripe with holes
that was introduced in Sec.~\ref{sec-diagstripe}.
Again we do not exhaustively enumerate all possible states with
a given number of particles $M$. Instead, we build basis states
from a starting state which has the stripes all merged together. 
The motivation of these studies is that understanding of 
the stripe-stripe repulsion is a prerequisite to
calculating the stripe-array energy as a function
of filling near $n=1/2$, which in turn is needed in 
order to ascertain the density at which sthe
stripe array would coexist with a liquid phase. 
Unfortunately, we can only study comparatively short stripes 
and the asymptotic form for $L_x \to \infty$ is quite unsure
from our numerical results.

\subsection{Boson and fermion statistics}
\label{sec-2andmoreBoseFermi}

\begin{figure}[ht]
	\centering
	\includegraphics[width=0.6\linewidth]{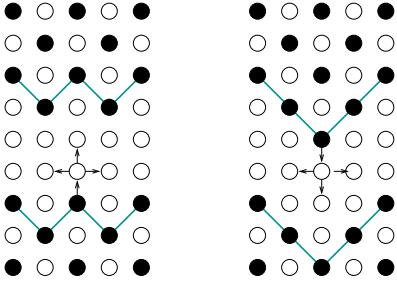}
	\caption{
	Two configurations of two 
	merged stripes. The arrows indicate the possible
	hops of one particle along the stripe. The horizontal movement 
	is limited to the adjacent columns only. 
	}
	\label{fig-2stripehops}
	\end{figure}

We observe from exact diagonalization that
for rectangular lattices with
two stripes, the boson and fermion spectra
are identical.\cite{FN-close}
In Fig.~\ref{fig-2stripehops}, we show two configurations
of two merged stripes. For a particle on the stripe we see
that the horizontal hops are limited by
nearest-neighbor replusion to adjacent columns only.
This means that, with two stripes, particles cannot move freely
along the stripe and the hopping in the system is still
primarily in the vertical direction. From Fig.~\ref{fig-2stripehops},
we can also see, after trying some moves,
that we cannot move one particle far away enough so as to exchange
the position of two particles. This is a necessary condition
for the boson and fermion spectra are 
identical.

We also need to show that periodic boundary conditions
do not affect the spectra, and we use again the idea outlined
in Sec.~\ref{sec-statistics}.
We have, as we did for the one-stripe-with-a-hole problem
in Sec.~\ref{sec-onehole},
numerically checked a number of systems with two stripes,
for example $4\times 10$, $6\times 8$, and $8\times 10$,
and we find that for the cases we checked 
the function $\sigma(\alpha)$ is well defined.

\subsection{Stripe-stripe interaction}
\label{sec-stripestripe}

\begin{figure}[ht]
	\centering
	\includegraphics[width=0.8\linewidth]{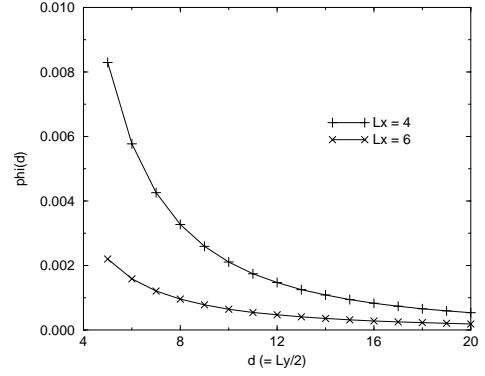}
	\includegraphics[width=0.8\linewidth]{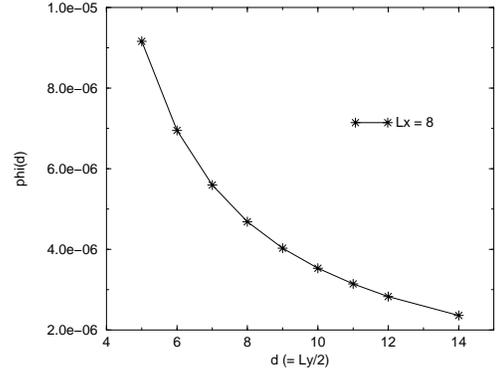}
	\caption{
	Stripe-stripe interaction energy cost $\phi(d)$ (Eq.~(\ref{eq-phi}))
	for stripe length 4, 6, and 8. $d$ is the distance
	between adjacent stripes ($d=L_y/2$).
	The lines are those that connect the data points.
	}
	\label{fig-phi}
	\end{figure}

>From diagonalization of two stripes with stripe length
$L_x=4,6,8$, we observe that the highest-weight states
in the ground state eigenvector have stripes far
apart from each other, which suggests
that stripes may repel. To study the interaction 
between two stripes quantitatively, we define
the following function,
	\begin{equation}
	\phi^{L_x}(d)=\frac{E_{2stripes}(L_x,L_y)-2 E_{stripe}(L_x)}{2 L_x},
	\label{eq-phi}
	\end{equation}
where $d$ is the distance between adjacent stripes 
(here $d=L_y/2$ for two evenly spaced stripes),
$E_{2stripes}$ the energy of the two-stripe system, and
$E_{stripe}$ the energy of one single stripe.
$\phi$ is the energy cost per unit length
per stripe due to stripe interaction,
and it is positive for repelling stripes
and negative for attracting stripes.
(When we are not considering the dependence of 
$\phi^{L_x}(d)$ on $L_x$, we will simply use $\phi(d)$.)
In Fig.~\ref{fig-phi} we show
$\phi(d)$ for stripe length 4, 6, and 8.
For all cases, $\phi(d)$ is positive (the stripes repel)
and decays as the stripe separation $d$ increases.
Observe that $\phi(d)$ for $L_x=8$
is very much smaller than that for $L_x=4,6$. 
The behavior of $L_x=8$, which is not yet explained, is highly significant for
extrapolation to the thermodynamic limit. 
The stripes with $L_x=4,6$ were so short that, from
Sec.~\ref{sec-onestripe} up to here, 
their $L_y$ size dependence could be explained in
terms of a particle moving in one dimension ($y$), 
completely ignoring their internal fluctuations. 

\begin{figure}[ht]
	\centering
	\includegraphics[width=\linewidth]{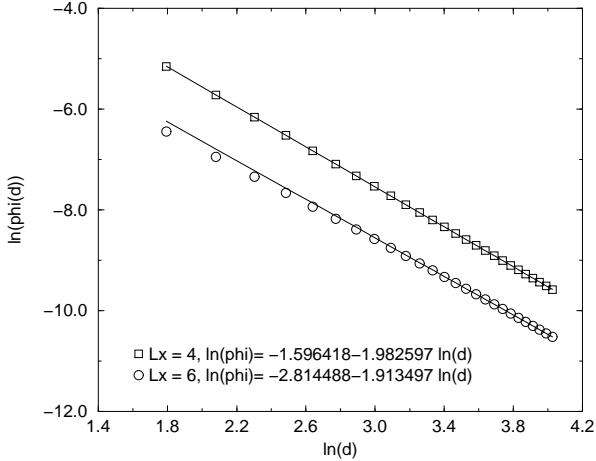}
	\caption{
	log-log plot of energy cost function
	$\phi(d)$ (Eq.~(\ref{eq-phi})) for stripe length $L_x=4,6$,
	with $d=6,8,...,56$.
	The slope (decay exponent $\alpha$ in Eq.~(\ref{eq-alpha})) 
	is close to 2 for $L_x=4$,
	and for $L_x=6$ the slope approaches 2 as stripe separation $d$
	increases. Linear regression is used for all $L_x=4$ data,
	and for the $L_x=6$ data, only the ten points with the largest $d$
	are used.
	}
	\label{fig-philnln}
	\end{figure}

It is clear that as $d\rightarrow \infty$,
$\phi(d)\rightarrow 0$ and from the graphs it does not
decay exponentially. We try the following 
power-law fitting function,
	\begin{equation}
	\phi(d)=\frac{A(L_x)}{d^\alpha}.
	\label{eq-alpha}
	\end{equation}
In Fig.~\ref{fig-philnln} we plot $\ln(\phi(d))$ vs $\ln(d)$
for $L_x=4,6$ with $d=6,8,...,56$, i.e., 
the largest lattice for $L_x=4$ is $4\times 112$
and for $L_x=6$ is $6\times 112$.
We see that the power-law assumption is good for $L_x=4$
with the decay exponent $\alpha$ (slope in Fig.~\ref{fig-philnln})
close to 2. For $L_x=6$ we observe that the 
slope approaches 2 as the stripe separation 
$d$ is large.

\subsection{Power-law decay and stripe effective mass}
\label{sec-mass}

In Sec.~\ref{sec-doublewell}, we explained the exponential
decay of the one-hole-with-a-stripe energy in $L_y$
by mapping the stripe motion to a one-dimensional problem 
with a double-well potential. Here with two stripes in the
system, we have shown again that stripe fluctuations as a function
of $x$ are limited, and we expect a one-dimensional potential
can be sufficient in capturing the essential physics.

Here we consider the two stripes as hard-core particles of
mass $m^*$ moving in the $y$ direction only.
This can be mapped to a pair of fermions on a one-dimensional
ring of length $L_y$, mathematically just like the 
one-dimensional fermions 
of Sec.~\ref{sec-1stripephase}, 
that move in the $x$ direction, 
but with a completely different relation to the physics.~\cite{FN-reducedmass}
The wavevectors of these one-dimensional fermions 
are thus $\pm \pi/L_y$, so the total energy is
	\begin{equation}
	E=\frac{\hbar^2}{2m^*}2 \left(\frac{\pi}{L_y}\right)^2
	=\frac{\hbar^2 \pi^2}{m^* L_y^2}.
	\label{eq-box}
	\end{equation}

For the two-stripe problem, from curve fitting in 
Fig.~\ref{fig-philnln}, the decay exponent
$\alpha$ is close to 2 for $L_x=4$. Using 
Eq.~(\ref{eq-alpha}) for $\phi(d)$, the definition 
for $\phi(d)$ in Eq.~(\ref{eq-phi}), and $d=L_y/2$ 
for two evenly spaced stripes, we have the following
formula for the two-stripe interaction energy  
	\begin{equation}
	E_{2stripes}-2E_{stripe}=\frac{8L_x A}{L_y^2},
	\label{eq-inter}
	\end{equation}
where $A$ is the factor in Eq.~(\ref{eq-alpha}).

Eqs.~(\ref{eq-box}) and (\ref{eq-inter}) give an expression for
the effective mass of a stripe from
two-stripe interaction,
	\begin{equation}
	m^*(L_x)=\frac{\pi^2}{8L_x A}.
	\label{eq-mstar}
	\end{equation}
The linear fitting intercept in Fig.~\ref{fig-philnln} gives,
for $L_x=4$, $A=\exp(-1.596418)$, and 
Eq.~(\ref{eq-mstar}) then gives us the effective mass
$m^*(4)=1.5222$. In Table~\ref{t-m},
the $L_x=4$ stripe effective mass calculated 
in the one-hole-with-a-stripe problem is
1.3561, and in Table~\ref{t-m2} that from
the two-hole problem is 1.3417 for fermions
and 1.3358 for bosons. All these results 
for the effective mass of a short, length-four
stripe are comparable.

It is clear that the shorter the stripes the better
the one-dimensional approximating model is. 
For $L_x=6$, it can be seen in Fig.~\ref{fig-philnln} that
the exponent approaches 2 in the large-$d$ limit,
but for $d$ close to 56 (the largest
system that we calculated), linear regression still 
gives 1.91. The intercept for $L_x=6$ in Fig.~\ref{fig-philnln}
is not yet sufficient for us to use
Eq.~(\ref{eq-mstar}) to calculate the 
stripe effective mass.

\subsection{Three and four stripes}
\label{sec-3and4stripe}

\begin{figure}[ht]
	\centering
	\includegraphics[width=0.75\linewidth]{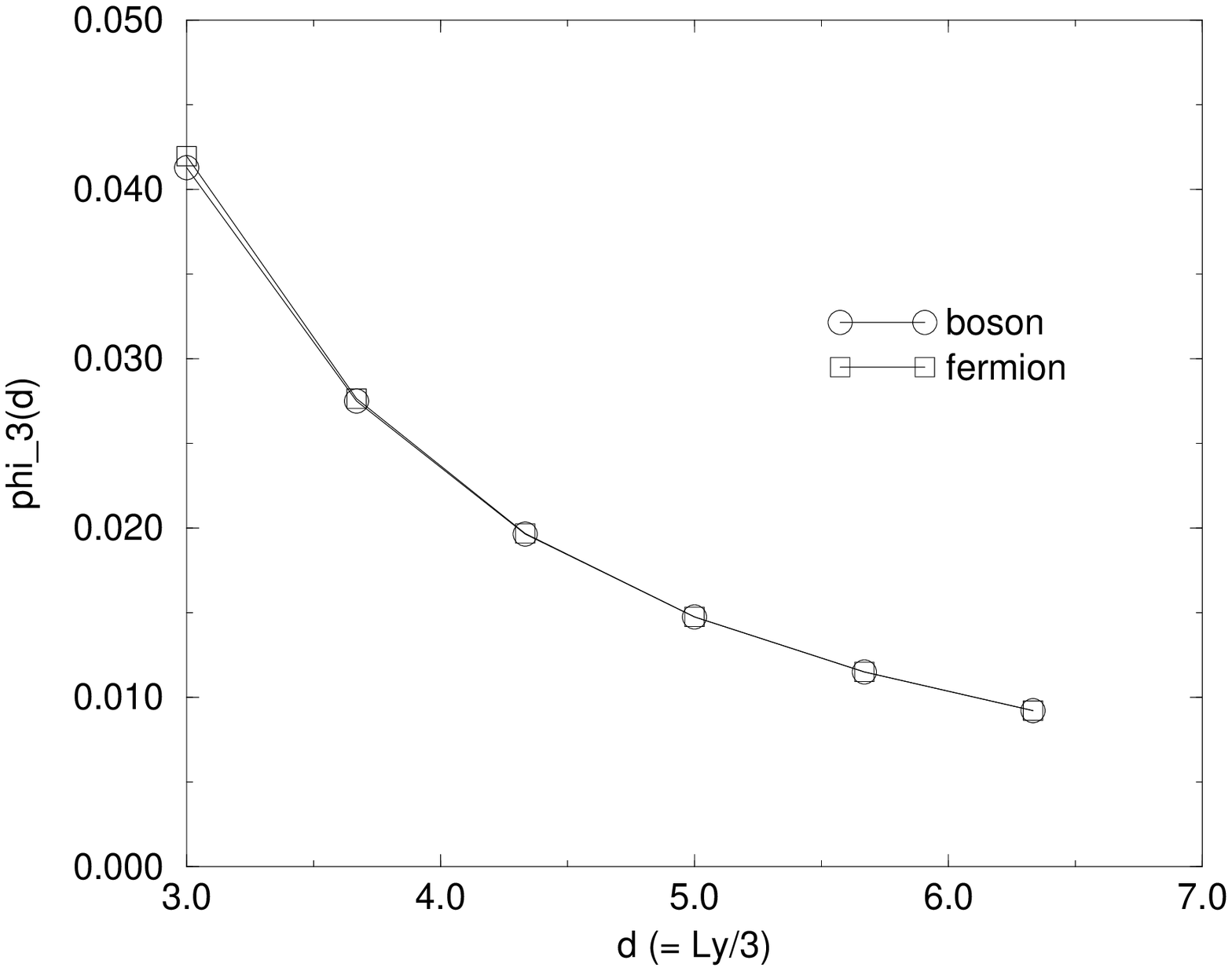}
	\includegraphics[width=0.75\linewidth]{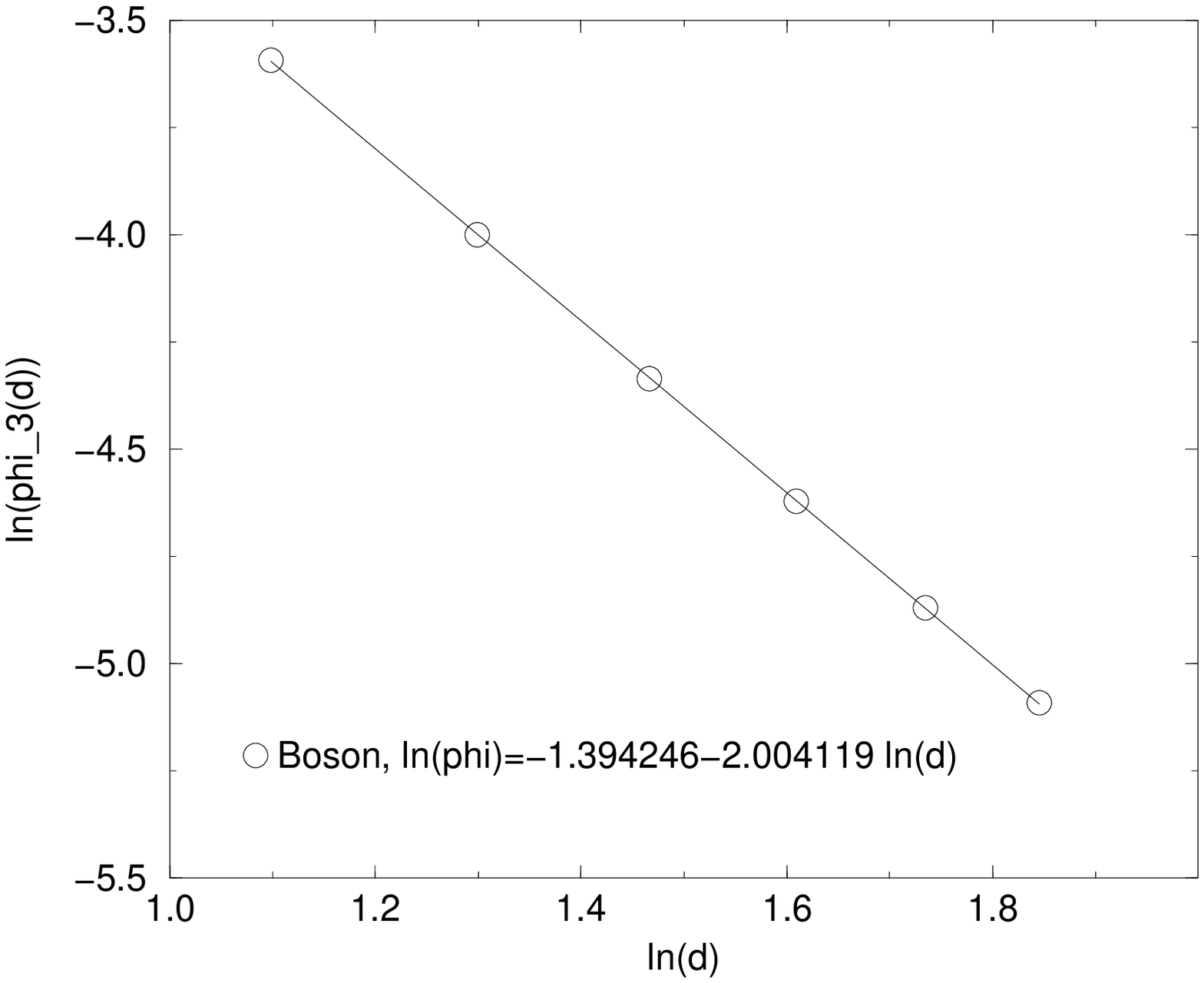}
	\caption{
	Stripe-stripe interaction energy cost $\phi_3(d)$ (Eq.~(\ref{eq-phi3}))
	for three stripes with length 4 ($L_x=4$). $d$ is the distance
	between adjacent stripes ($d=L_y/3$). The boson 
	and fermion energies are slightly different.
	The log-log plot is shown below, showing
	for bosons an exponent $2.004$. 
	}
	\label{fig-phi3}
	\end{figure}

We have also studied the interaction of three
and four stripes with stripe length $L_x=4$. From 
diagonalization, we find that for 2, 3, and 
4-stripe fermion systems we computed, 
the ground state energy always appears in the $\kk=(0,0)$
sector, and the highest-weight states in the ground state
eigenvector have stripes far apart.
To study the interaction of three stripes,
we define, in the same fashion as Eq.~(\ref{eq-phi}), the energy cost
per unit length per stripe,
	\begin{equation}
	\phi_3(d)=\frac{E_{3stripes}(L_x,L_y)-3 E_{stripe}(L_x)}{3 L_x},
	\label{eq-phi3}
	\end{equation}
where $d=L_y/3$ here. In Fig.~\ref{fig-phi3}
we plot $\phi_3(d)$ vs $d$ and $\ln(\phi_3(d))$ vs
$\ln(d)$. First of all, with three or more stripes,
unlike the two-stripe case, particles have enough room 
to exchange with each other when
the stripes merge, and the fermion and 
boson energies are no longer the same.\cite{FN-phi3}
(However, we see from the graph that the boson and
fermion energies are only slightly different.)
It is clear that the stripes repel
and the exponent of $\phi_3(d)$ for $L_x=4$ is
2.004 for the boson case (and practically the same
for the fermion case, not shown in the graph), close to 1.983, the exponent for
the two-stripe $L_x=4$ case (see Fig.~\ref{fig-philnln}).

The stripes are short enough in the $x$ direction to be treated as rigid
(See Sec.~\ref{sec-onestripe}). 
They are, in effect, three hardcore particles moving in one
dimension. 
Apart from the fact that this one dimension runs along $y$, 
exactly the same problem was studied in Sec.~\ref{sec-1Dfermion}, 
where we recalled that it is convenient to convert such particles to fermions. 

Let us consider the ground state of three fermions of mass $m^*$
in a length-$L_y$ box. They have wavevectors $q_y=0$ and $\pm 2\pi/L_y$, 
so the total energy is
	\begin{equation}
	E=\frac{\hbar^2}{2m^*}\left(0+2\left(\frac{2\pi}{L_y}\right)^2\right)
	=\frac{4\pi^2/m^*}{L_y^2}.
	\label{eq-box3}
	\end{equation}
For the three-stripe problem, we use the same $1/d^2$ relation
for $\phi_3(d)$, as in Eq.~(\ref{eq-alpha}) for the two-stripe problem,
	\begin{equation}
	\phi_3(d)=\frac{A_3(L_x)}{d^2},
	\label{eq-alpha3}
	\end{equation}
where $A_3(L_x)$ is a constant. Rearranging
Eqs.~(\ref{eq-box3}) and (\ref{eq-alpha3}) gives us a formula
for stripe effective mass $m^*(L_x)$, 
similar to the two-stripe formula in Eq.~(\ref{eq-mstar}),
	\begin{equation}
	m^*(L_x)=\frac{4\pi^2}{27 L_x A_3(L_x)}.
	\label{eq-mstar3}
	\end{equation}
Using the linear fitting intercept in Fig.~\ref{fig-phi3},
we get for $L_x=4$, $A_3(4)=\exp(-1.394246)$, and
then from Eq.~(\ref{eq-mstar3}), we get $m^*(4)=1.4738$,
which is consistent with the two-stripe result
$1.5222$ calculated in Sec.~\ref{sec-mass}.

We have also calculated the energy for the smallest
system with four stripes: the $4\times 12$ system with
16 particles. Here the boson energy
is $-10.8525$ and the fermion energy $-10.8418$, both
higher than the energy of four indepedent stripes with
length 4, $-11.3137$.

\section{The Stripe-array}
\label{sec-stripearray}

In this paper, we have studied the limit near the half-filled,
checkerboard state, with cases of a single stripe, holes-on-a-stripe,
and a few stripes. As we approach intermediate fillings,
around 1/4, we could have two situations: the separation of 
the system into particle-rich and hole-rich regions
(sometimes argued to be quite general for interacting 
fermions with short-range interactions~\cite{visscher,Emery90})
or else an array of stripes. How do we know which state is the 
ground state for our model? How do we know whether 
the stripe-array state is stable?

If the true behavior, in the thermodynamic limit, 
is phase separation between the half-filled checkerboard
state and a hole-rich liquid, then -- in a sufficiently large
finite system -- the ground state would consist of one large
(but still immobile) droplet in a checkerboard background, 
of the sort explained in Subsec.~\ref{sec-droplets}. 
However, in the smallish systems tractable by exact diagonalization, 
the ground state would probably be a stripe-array state, for
(as we find in Sec.~\ref{sec-stability}), 
the bulk energy density difference is quite small between the
stripe-array state and the phase-separated state, 
yet the latter state must pay a considerable extra cost of the
phase-boundary line tension times the droplet's perimeter. 

On the other hand, the stripe-array state in a finite system is
extremely sensitive to boundary conditions: the concentration of
holes may be just right to form one stripe across the system, but
if the boundary vectors are both even, the system can 
only support an even number of stripes. Even if the 
stripe-array is the correct thermodynamic answer at this
hole filling, the best state available to this finite system is
a separated droplet.  

We conclude that one cannot accept the behavior of our finite
systems as a direct picture of the thermodynamic limit, because
finite systems are dominated by finite-size and topological effects.
The correct approach is to determine the
equation of state for each competing phase by extrapolating
it to the thermodynamic limit, and only then to perform 
a Maxwell construction to determine the phase stability. 
Note that different extrapolation schemes may be
appropriate for qualitatively different phases.

\subsection{Stripe-array chemical potential}
\label{sec-chem}
We first calculate 
the chemical potential of the stripe-array, i.e., the
energy per hole in creating a new stripe;
this information will be needed for stability analysis later.
We have previously presented this calculation in a condensed
form.\cite{HenleyZhang}

Say we have a $L_x\times L_y$ system with $p$ stripes 
stretching in the $x$ direction. The number of particles per
column is $(L_y-p)/2$ and the total number of particles
is $M=(L_y-p)L_x/2$. Denote the total number 
of lattice sites $N=L_x L_y$, then  the particle density is 
	\begin{equation}
	n=\frac{M}{N}=\frac{1}{2}-\frac{p}{2L_y}=\frac{1}{2}-\frac{1}{2d},
	\end{equation}
where $d$ is the separation between adjacent stripes
and for even distributed stripe we have $d=L_y/p$.
We then have
	$d=1/(1-2n)$.

The energy of the stripe-array has two contributions:
the energy from indepedent stripes and the energy due to 
stripe-stripe interaction. The energy per length $\sigma_0$
of an infinitely long stripe is 
$\sigma_0=-2/\pi$,
which can be obtained from mapping the stripe to the 
spin-1/2 chain.\cite{Mila}
Therefore the energy of $p$ independent stripes is 
	\begin{equation}
	E_{indep}=p\,\sigma_0 L_x=\frac{\sigma_0}{d} L_x L_y,
	\label{eq-indep}
	\end{equation}
where we have used $p=L_y/d$. On the other hand,
the energy due to stripe-stripe interaction is
	\begin{equation}
	E_{interaction}=p\,\phi(d) L_x=\frac{\phi(d)}{d} L_x L_y,
	\label{eq-interaction}
	\end{equation}
where we have used the two-stripe interaction energy per length
per stripe $\phi(d)$ defined in Eq.~(\ref{eq-phi}).
(Note that we have seen in Sec.~\ref{sec-stripestripe}
that $\phi$ also depends on the length of the stripe
$L_x$. Fortunately, in the following, we will only work in
the $d\rightarrow \infty$ limit of $\phi(d)$, which
is independent of $L_x$.)
Combining Eq.~(\ref{eq-indep}) and Eq.~(\ref{eq-interaction}),
we obtain the energy density of the stripe array
as a function of particle density $n$,
	\begin{eqnarray}
	\nonumber
	&&{\cal E}_{sa}(n)=\frac{E_{indep}+E_{interaction}}{N}\\
	&&=(\sigma_0+\phi(d))\frac{1}{d}=(\sigma_0+\phi(d))(1-2n).
	\end{eqnarray}
Furthermore, we have seen, when we studied stripe-stripe 
interaction, $\phi(d)$ is positive and $\phi(d)\rightarrow 0$
as $d\rightarrow \infty$. So in the infinite-lattice limit
we have
	\begin{equation}
	{\cal E}_{sa}(n)=\sigma_0(1-2n).
	\end{equation}
That is to say that the chemical potential of the stripe-array
is 
	\begin{equation}
	\mu^*=\frac{d{\cal E}_{sa}(n)}{dn}=-2\sigma_0=\frac{4}{\pi}
	=1.273,
	\label{eq-mu}
	\end{equation}
where the result for $\sigma_0$ 
is used. Eq.~(\ref{eq-mu}) says that the slope of the 
stripe energy density curve is $1.273$,
and if we have an array of stripes in the system 
then adding a particle to the system will raise
the energy by $1.273$, or equivalently, adding a hole to
the stripe-array will lower the energy by $1.273$.
This is an important quantity in the stability analysis
that is to follow.

\subsection{Stripe-array stability}
\label{sec-stability}

We consider three stability conditions for the stripe-array
agains phase separation.

\subsubsection{Stability condition 1: stripes repel}

We have already studied stripe-stripe interaction,
and we know that stripes repel (see Fig.~\ref{fig-phi}
for example). This is our first
stability condition.

\subsubsection{Stability condition 2: two stripes beat one fat stripe}

Our second stability condition comes from
the work on the one-hole-with-a-stripe problem in Sec.~\ref{sec-oneholeLx}.
Here we want to know whether with more holes
new stripes will form or the existing stripes just get 
wider (thus leading to phase separation).
We showed in Sec.~\ref{sec-oneholeLx}
that adding a hole to an infinitely
long stripe lowers the energy by 0.66.
Here we have shown that adding a hole to a stripe-array
lowers the energy by 1.273. The stripe-array
state is therefore preferred.

\subsubsection{Stability condition 3: stripe-array beats phase separation}

\begin{figure}[ht]
	\centering
	\includegraphics[width=\linewidth]{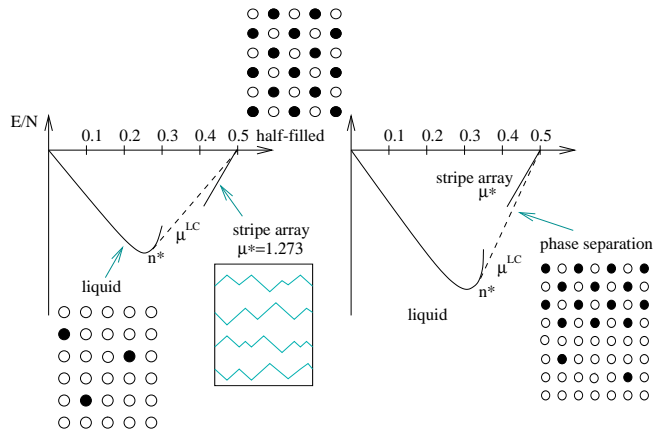}
	\caption{
	Stability of the stripe-array vs phase separation.
	Energy density $E/N$ vs particle density $n=M/N$.
	The stripe-array curve has slope $\mu^*=1.273$.
	The dashed line is tangent to the 
	liquid curve and is connected to the half-filled state.
	It represents the coexistence of the two phases: liquid
	and half-filled states, i.e., a phase separated state.
	On the left is the case where the stripe-array line
	is below the dashed line, and therefore the stripe-array
	is stable. On the right is the case where the dashed line
	is below the stripe-array line, i.e., the phase separated
	state is stable. $n^*$ is the intersection
	of the tie-line with the liquid curve, and 
	the key quantity is the slope
	of the dashed line $\mu^{\rm LC}$.
	}
	\label{fig-stability}
	\end{figure}

The third stability condition comes
from free energy analysis in statistical mechanics.
Here because we are considering zero temperature
physics and $F=E-TS$, energy is free energy.
In Fig.~\ref{fig-stability} we show the two cases.
The dashed tie-line is the Maxwell construction. 
It is tangent to the liquid curve (small fillings) 
and is connected to the half-filled
state at $n=1/2$. It represents the coexistence of 
the liquid state and the half-filled state,
i.e., the phase separated state. According to
Sec.~\ref{sec-chem}, the stripe-array line should have slope
1.273, and it is drawn from the half-filled limit.
The stripe-array case is stable when the stripe-array line
is below the dashed line, otherwise the phase separated state
is stable. Therefore, to determine the stability
of the stripe-array against phase separation, we need
to determine the slope of the dashed line
$\mu^{\rm LC}$. Here we are following the notation
in Ref.~\onlinecite{HenleyZhang} and the superscript LC denotes the two states:
liquid and CDW (charge-density-wave, i.e., half-filled state)
that are connected by the dashed line.

To obtain $\mu^{\rm LC}$, we need a fitting function
for the energy density $E/N$ at intermediate fillings (the liquid part).
We use a polynomial form, up to the third order
in particle density $n=M/N$,
	\begin{equation}
	{\cal E}(n)=\frac{E(n)}{N}=A_1 n + A_2 n^2 + A_3 n^3,
	\label{eq-fit}
	\end{equation}
where $E(n)$ is the energy of the $M$ particle system.
We should emphasize that this fitting form Eq.~(\ref{eq-fit})
is not meant for the dilute limit ($n\rightarrow \infty$).
This is for the intermediate-filling, with $n\approx 1/4$.
Eq.~(\ref{eq-fit}) can also be written as
	\begin{equation}
	\frac{E}{M}=A_1 + A_2 n + A_3 n^2,
	\label{eq-fitM}
	\end{equation}
where $E/M$ is the energy per particle. We will use
Eq.~(\ref{eq-fitM}) to fit diagonalization results.
The slope $\mu^{\rm LC}$ can be determined by first solving
for $n^*$, the $n$ coordinate of the intersection of the dashed tie-line
with the liquid curve, using
	\begin{equation}
	\mu^{\rm LC}=
	\frac{{\cal E}(n^*)-0}{n^*-1/2}=\frac{d{\cal E}(n)}{dn}\Bigg|_{n^*}.
	\label{eq-muLC}
	\end{equation}

	\begin{figure}[ht]
	\centering
	\includegraphics[width=0.8\linewidth]{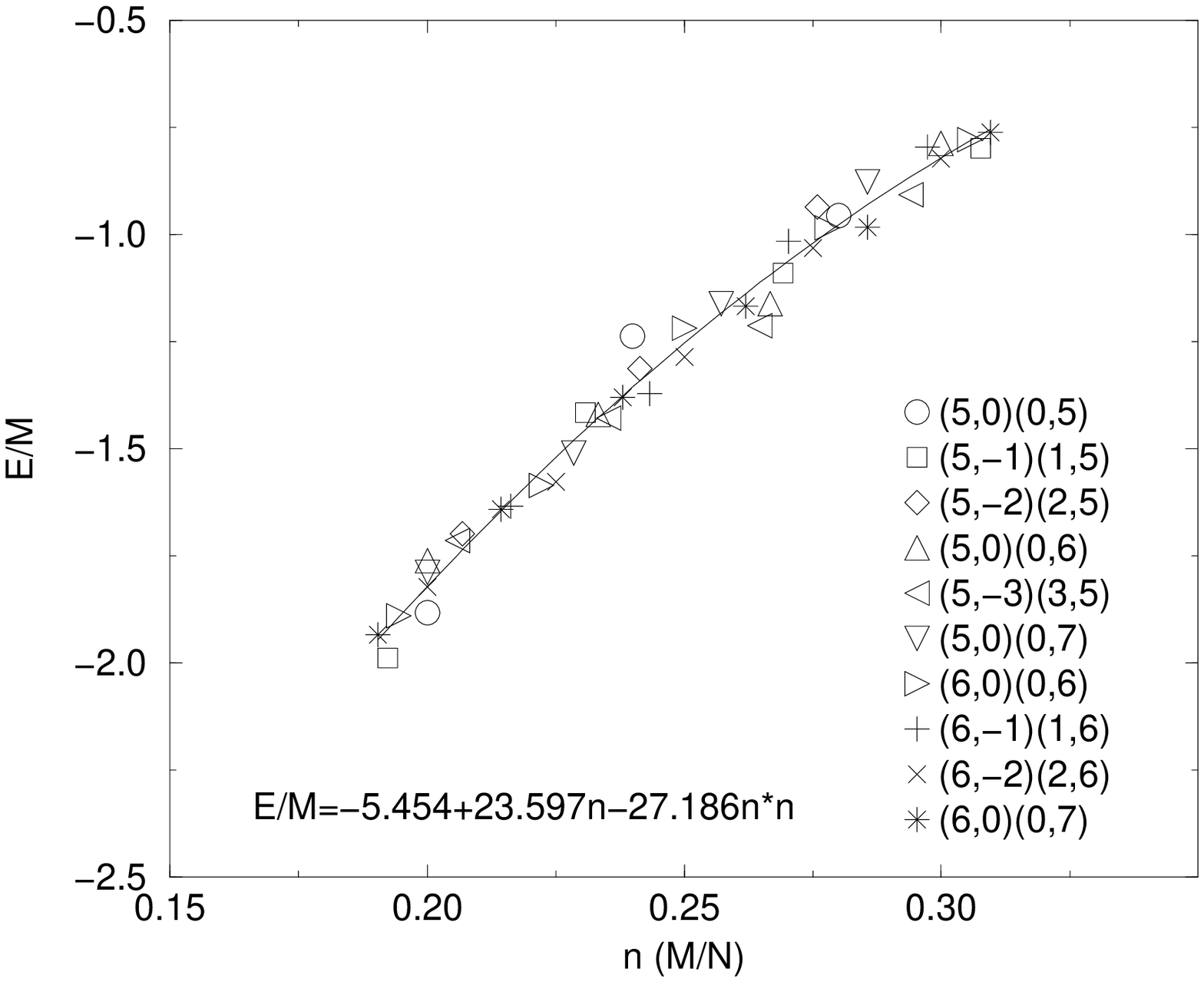}
	\includegraphics[width=0.8\linewidth]{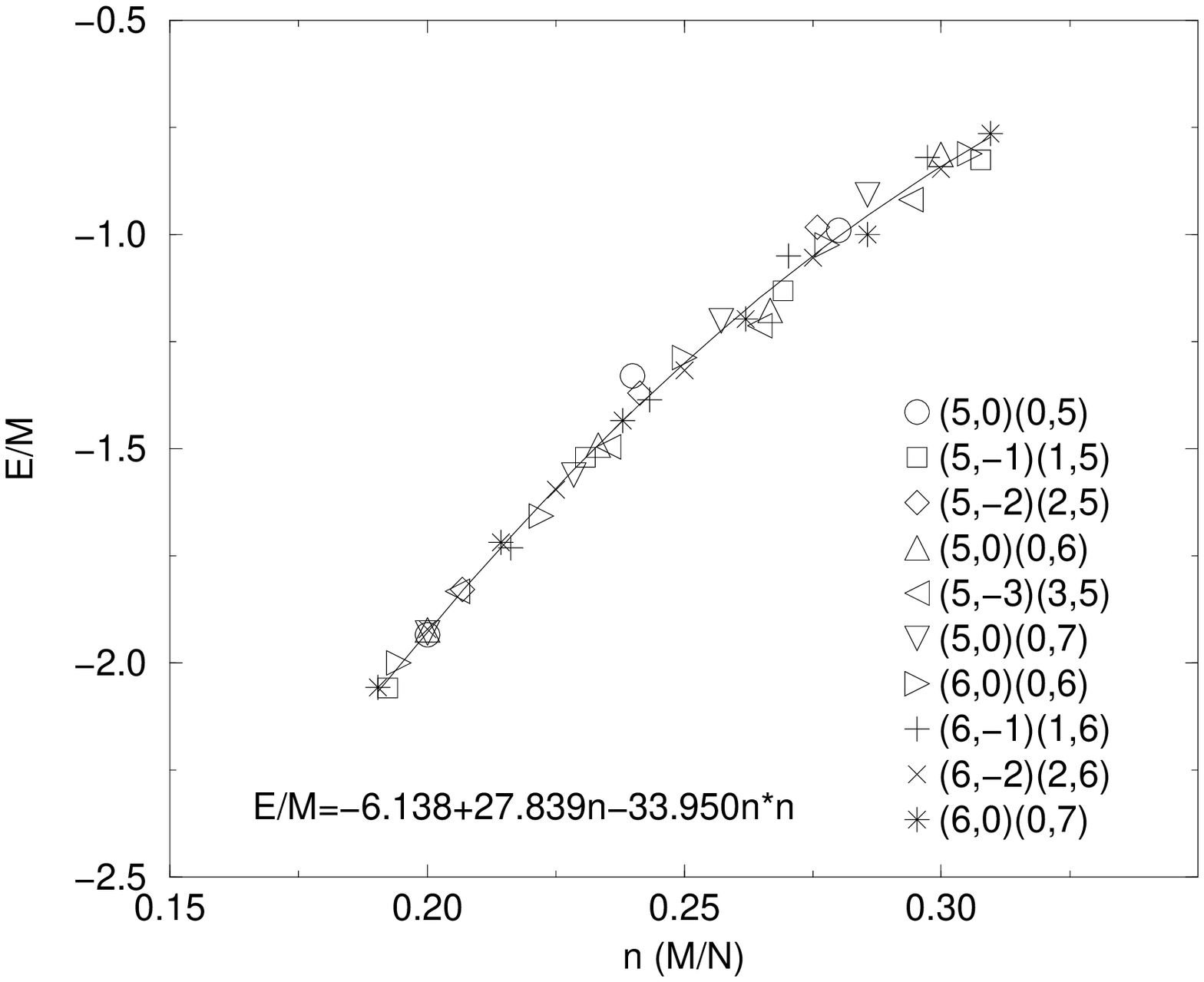}
	\caption{
	Fermion (top) and boson (bottom) $E/M$ vs $n=M/N$ fit
	for intermediate fillings ($0.19\le n \le 0.31$).
	The fermion data have larger spread than the boson
	data because of the fermion shell effect 
	}
	\label{fig-fit}
	\end{figure}

In Fig.~\ref{fig-fit}, we plot, for fermions and bosons, $E/M$ vs $n$ at
the intermediate fillings ($0.19\le n \le 0.31$)
for a number of lattices with lattice size
ranging from 25 to 42.\cite{FN-lattices}
In Table~\ref{t-stable} we list the quadratic
fitting parameters obtained from diagonalization data.
For the ten lattices in Fig.~\ref{fig-fit},
using Eq.~(\ref{eq-muLC}), we obtain
$\muLC=1.254$ for fermions
and $\muLC=1.304$ for bosons.

	\begin{table}[ht]
        \centering
        \caption{
	Fermion and boson fitting parameters (in Eq.~(\ref{eq-fit}))
	and the calculated $n^*$ and $\muLC$ (using Eq.~(\ref{eq-muLC})),
	for all ten lattices in Fig.~\ref{fig-fit} and for the 
	smallest and largest five lattices separately.
	Compared with the stripe-array slope $\mu^*=1.273$,
	the fermion stripe-array is stable against
	phase separation ($\muLC<\mu^*$),
	and the boson stripe-array is unstable ($\muLC>\mu^*$).
	}
        \begin{tabular}{c|c|ccc|cc|c}
	Particle & Lattice & $A_1$ & $A_2$ & $A_3$ & $n^*$ & $\muLC$ & SA Stable?\\\hline\hline
	Fermion & all 10 & -5.454 & 23.597 & -27.186 & 0.251 & 1.254 & Stable \\
	& small 5 & -5.995 & 28.230 & -36.825 & 0.255 & 1.240 &  \\
	& large 5 & -5.102 & 20.586 & -20.932 & 0.250 & 1.264 &  \\\hline
	Boson & all 10 & -6.138 & 27.839 & -33.950 & 0.233 & 1.304 & Unstable \\
	& small 5 & -6.378 & 29.929 & -38.420 & 0.231 & 1.301 & \\
	& large 5 & -5.994 & 26.580 & -31.251 & 0.233 & 1.307 & \\
        \end{tabular}
        \label{t-stable}
	\end{table}

In Table~\ref{t-stable}, we have also included calculations
for the five smallest and five largest lattices in Fig.~\ref{fig-fit}
separately. We find that $\muLC$ for both boson and fermion
increases with increasing lattice size. For the boson
case, this means that for large lattices, the stripe-array
is not stable against phase separation ($\muLC>\mu^*=1.273$). 
For the fermion case, the result is stable ($\muLC<\mu^*$)
for the lattices that we have studied, but
$\muLC$ is very close to the stability limit 1.273. 
The fermion stripe-array is possibly stable.

In our earlier publication, Ref.~\onlinecite{HenleyZhang}, 
we used the same polynomial fitting function,
Eq.~(\ref{eq-fit}), but we fixed $A_1=-4$, corresponding
to the energy of an noninteracting particle,
and our systems there went from 20 sites ($4\times 5$)
to 36 sites ($6\times 6$).
The analysis in this section
and the results in Table~\ref{t-stable} are obtained
using the three-parameter ($A_1,A_2,A_3$) fit
for bigger lattices, from 25 sites ($5\times 5$)
to 42 sites ($6\times 7$).
The conclusion in Ref.~\onlinecite{HenleyZhang} was
$\muLC=1.25(2)$ for fermions, 
and $\muLC=1.33(2)$ for bosons.

Ref.~\onlinecite{hebertbatrouni} simulated the boson case of our model
Eq.~(\ref{eq-Ham}), using Quantum Monte Carlo, but with $V<\infty$. 
Their phase diagram, Fig.~11 of Ref.~\onlinecite{hebertbatrouni}, shows
a first-order phase transition between the $(\pi,\pi)$ solid
[i.e. the checkerboard phase] and a superfluid [the hole-rich
phase with $n\leq 0.25$.  Our results correspond to a single point
$(t/V,\mu/V)=(0,0)$ on their plot; however, their phase boundary
approaches that point with a slope~\cite{batrouni-private} $2.28$, i.e. at 
$\mu/t = 2.28$ in the $V\to \infty$ limit, or $\mu/t=2.4$ in a related
simulation.~\cite{troyer}.  This is much
{\it larger} than our result $\mu_{\rm LC}\approx 1.33$. 
That is puzzling, since it would imply either that the liquid state (with 
density $n\approx 0.25$) has much lower energy than we found, 
or else that a state exists at $n\approx 0.4$ (in these finite systems)
having much lower energy than the stripe states we studied. 

The evidence in Ref.~\onlinecite{hebertbatrouni}
for phase separation rests on histograms such as
their Fig.~10 which exhibits (for $V=2.86$ in an $8\times 8$ 
system) a bimodal probability
distribution for the density $n$, in a grand canonical ensemble, 
with peaks at $n=0.5$ (the checkerboard state) and $n\approx 0.39$, 
the latter being interpreted as a liquid state. 
For our $V=\infty$ model, a similar plot at low temperature
would show peaks at $n=0.5$ and at $n=0.375$. The latter peak 
represents a two-stripe state with 8 holes. 
Its energy is $\approx -8\sqrt{1 + 1/\sqrt 2}$ $\approx -10.45$
(using Eq.~(\ref{eq-stripeE0}), and neglecting the stripe-stripe
interaction in light of Fig.~\ref{fig-phi}). 
Our point is that examination of the most probable configurations
is needed in order to judge whether the coexisting state contains
stripes or liquid droplets.   Indeed, the discrete coexistence 
between $n=0.5$ and $n=0.375$ is an artifact of the periodic
boundary conditions -- in the thermodynamic limit, within the
stripe-array phase, density $n$ varies continuously with the
chemical potential.

\section{Conclusion}

\subsection{Summary}

In this paper we studied systematically a two
dimensional model of strongly-interacting
spinless fermions and hardcore bosons
on the square lattice Eq.~(\ref{eq-Ham})
near the half-filled limit.
We considered an extended, 
quantum-fluctuating object
that is natural in our model at 
the dense limit--the stripe. 
After introducing our model (Sec.~\ref{sec-intro}) and
our diagonalization
program (Sec.~\ref{sec-diagstripe}),
we made a detailed study of
the problems of a single stripe (Sec.~\ref{sec-onestripe}), 
up to two holes on a stripe 
(Secs.~\ref{sec-onehole} and \ref{sec-twoholes}),
stripe-stripe interaction of two, three, and four
stripes (Sec.~\ref{sec-twoandmore}), and finally 
the stability of an array of stripes 
(Sec.~\ref{sec-stripearray}). Our theoretical studies
were aided with substantial amount of results
obtained from exact diagonalization.

As is well-known, quantum systems
in two dimensions are much more 
difficult to study than those
in one dimension. Because of the presence
of strong repulsion, at the dense limit
of our model, motion is severely limited.
We showed that for the problems of 
a single stripe, holes on a 
stripe and a few stripes, analytical
results can be obtained by mapping the two-dimensional
problems to one-dimensioinal ones, which can in turn
be studied using the analytical tools available in
one dimension. We showed in great detail the intricacies
of each map, including for example, the many-to-one problem,
the resulting phase factors, and the role of boson
and fermion statistics.

Another theme in this paper is the comparison
between the boson and fermion problems.
At the center of this
comparison is the state graph introduced
in Sec.~\ref{sec-diagstripe} which describes
graphically the relationship among the basis 
states. For the problems of a single stripe,
holes on a stripe, and a few stripes, we investigated
the effect of particle statistics on eigenenergy,
with the help of the state graph and our computer
program. And for the problem of stripe-array stability, 
we obtain from diagonalization results an interesting conclusion
that the boson stripe-array is not stable and
the fermion stripe-array is very close to the stability
limit and is possibly stable.

This paper taken as a whole is an attempt
to understand stripes as macroscopic objects
arising from microscopic components (the
spinless fermion and hardcore boson particles
in our model). The underlying physics
is the quantum mechanics of the component
particles that is described by a many-particle
Schrodinger equation (which we solve by
exact diagonalization). Stripes can perhaps
be called an {\it emergent phenomenon} and are
results of collective motions of 
many microscopic particles (the
original particles or holes). They can
perhaps be considered in the same fashion
as phonons arising from collective lattice
vibrations. And like phonons, the stripes 
in our problem have a life of their own,
and we have tried, in this paper,
to understand the new physics they bring.

This paper is one of the first using
exact diagonalization to study stripes.
This is possible because
of the significant Hilbert space reduction
afforded by our model Eq.~(\ref{eq-Ham}),
spinless fermions and hardcore bosons
with infinite nearest-neighbor repulsion.
We believe that this is one of the simplest
microscopic lattice model with which to study
stripes from the underlying particle dynamics.
And as we commented in our other paper
on the dilute limit of this model,\cite{dilutepaper}
this model deserves to be better known 
and better studied.

\subsection{Future directions}

There are a number of directions in which the 
work of this paper might be extended. 
For example, our code handles quite arbitrary 
periodic boundary conditions, not necessarily
rectangular or square. Our key use of this feature
was to accurately measure the equation of state
for the strongly interacting liquid at density $n\approx 0.25$, 
as used in Sec.~\ref{sec-stability} to decide the
stripe-array stability.  By including a wide variety
of differently shaped fermion systems in Fig.~\ref{fig-fit}
(top plot), in effect we average out the ``noise''
(which is due to fermion shell effects~\cite{dilutepaper}). 
An alternative way to do the same thing, which 
we have not implemented in the present work, 
would be to impose a phase factor
across the boundary conditions and average over
all phases.~\cite{phasetwist}

A boundary condition could be used to 
investigate the 90$^\circ$ bend of a stripe, 
which was discussed in Sec.~\ref{sec-stripe90}.
These would force a diagonal stripe, 
as in Fig.~\ref{fig-45hole}, 
but additional holes would be added to the stripe which
are expected to condense and form a segment at right angles. 
That geometry would permit estimation of the
energy cost of the $90^\circ$ bend. 


This study has focused entirely on eigenenergies; 
since our diagonalizations provide  the wavefunctions too, 
we could have computed a variety of informative expectations, 
such as correlation functions. 
We did make qualitative observations using
snapshots of the high-weighted configurations. 
In our diagonalizations with two or more stripes
(Sec.~\ref{sec-3and4stripe}), 
which started from a state with stripes all merged, 
the highest-weight state in
the ground state eigenvector
has stripes far apart from each other, 
and nearly equally spaced, as expected from
the one-dimensional approximate wavefunctions 
of Sec.~\ref{sec-twoandmore}. 
Calculation of the probability distribution
for the stripe separation would provide a
quantitative test of those wavefunctions. 
Similar calculations
would give a more direct check of the exponential
decay of the hole probability as a function of
its distance from a stripe, as predicted in
Sec.~\ref{sec-doublewell2}; and would 
immediately reveal the attractive or repulsive
tendency of two holes on a stripe (Sec.~\ref{sec-stripe90}). 


We studied the case of {\it one} stripe with {\it two} holes;
the case of {\it two} stripes with {\it one} hole is also
worth investigating. 
The finite-size dependence of the extra energy due to the hole
would shed light on the processes by which the hole is transferred
between stripes when they touch. Such processes are critical 
to the transport properties of the stripe array, as considered
in Refs.~\onlinecite{anisotransport} in the spinfull case. 

Up to here, we discussed further kinds of measurements
which could be made on the same model system;
variations are also possible of the model itself.
The most obvious of these is to treat $V<\infty$. 
A large but finite $V$ may be handled in the spirit of the $t$-$J$:
the same highly restricted basis states 
(which facilitated our diagonalizations)
are retained, but new hopping terms along [2,0] and [1,1] type
vectors appear,  with amplitudes $-t^2/V$, wherever the intervening
site would be forbidden by the nearest-neighbor exclusion. 
It is not obvious whether this tends to stabilize or to
destabilize the stripe array.  The inclusion of another
form of correlated hopping (Ref.~\onlinecite{HenleyZhang}, 
Fig.~3(c)) can certainly stabilize or destabilize the stripe
array depending on its sign.~\cite{HenleyZhang}

Finally, the same code is adaptable, almost without modifications,
to the triangular lattice. That may model 
$^3$He or $^4$He atoms adsorbed on graphite or
on carbon nanotubes,~\cite{green00,green02}
which implement a periodic boundary 
condition in one direction.

\acknowledgments
We acknowledge support by the National Science Foundation under
grant DMR-9981744. 
We thank G.~G.~Batrouni for helpful discussions.

\appendix 

\section {Counting basis states for the program}

\def \Nbas {{\cal N}}   
\def \Nbasstripe{{\Nbas^{\rm stripe}}}

We find the number of basis states $\Nbas$ is maximum for
fillings $n\equiv M/N \approx 0.25$, close to the filling $n^*$
at which, in the thermodynamic limit,
(see Sec.~\ref{sec-stability}) a hole-rich liquid coexists with the
half-filled state or the (nearly-half-filled) stripe array state. 
This count may be estimated by adapting Pauling's trick
for the ice model entropy~\cite{Pauling}. 
The number of ways merely to distribute $M$ particles over $N$
sites, uncontrained,  is ${N\choose {M}}$ implying the usual entropy
$-n \ln n - (1-n) \ln (1-n)$ as $N\to \infty$; this must
be corrected to account for the constraint of no nearest neighbors. 
One chosen particle has
a probability $(1-n)^4$ to be free of nearest neighbors. 
If we pretend this event is independent as each particle
on (say) the even lattice is chosen in turn, 
then the $M/2$ power of this probability is the chance 
for the whole configuration to be valid.
The net entropy is estimated as 
$ -n \ln n - (1-3n) \ln (1-n)$;
this attains its maximum $0.42$ around $n=0.24$. 
implying the leading dependence $\Nbas \sim 1.54^N$ for the
$V=\infty$ spinless case, compared to $\Nbas \sim 4^N$ 
in the Hubbard model case. 
In fact, for a $7\times 7$ lattice, the block with 
$M=11$ particles has the largest (translation-reduced) matrix
dimension, $\Nbas = 1 906 532$.

Let $\Nbas^s_h(L_x,L_y)$ be the
number of basis states (per ${\bf k}$ vector) 
in a configuration with $s$ stripes and $h$ holes on it. 
This grows exponentially with $L_x$
but comparatively slowly with $L_y$, 
so we can handle quite large $L_y$, 
and moderately large $L_x$. 

In the case of one stripe, the map to an 
XX spin chain (Sec.~\ref{sec-maptospin})
makes clear that, independent of $L_y$,
the basis has 
$\sim {L_x}^{-1} {{L_x}\choose{L_x/2}}$ states
(where ${A \choose B} \equiv A!/[B! (A-B)!]$). 
Asymptotically 
$\Nbas^1_0 (L_x,L_y)  \sim 2^{L_x}/{L_x}^{3/2}$. 

The number of states with two stripes
may be estimated by placing them independently -- the
correction from disallowing overlaps is subdominant. 
That is the square of the one-stripe count, 
multiplied by $L_xL_y$ for the possible vectors
offsetting one stripe relative to the other, i.e.
$\Nbas^2_0(L_x,L_y)  \sim 2^{L_x}/{L_x}^{3/2}$. 
$\sim 4^{L_x} L_y/L_x^2$.

For one stripe with one hole, 
the basis states were counted for $L_y\leq 25$ and $L_x\leq 10$.
Empirically, 
    \begin{equation}
      \Nbas^1_1(L_x,L_y) = \frac{1}{2}  
      {{L_x}\choose{L_x/2}} (L_y-1) +B(L_x)
    \label{eq-count11}
    \end{equation}
exactly, where $B(4)=-2$, $B(6)=-4$, $B(8)=-8$, $B(10)=-14$; for $L_y=7$, this
appears to follow with $B(12)=-12$, $B(14)=66$, $B(16)=572$. 
The linear increase with $L_y$ is due to the states with the hole
away from the stripe. 

For a stripe with two holes, 
Eq.~(\ref{eq-count11}) gets multiplied by an 
additional factor $L_xL_y$, 
for the possible offset of the second hole relative 
to the first.  Empirically, 
$\Nbas^1_2 (6,L_y) = 15{L_y}^2 - 47 L_y + 4$
(valid for $L_y>L_x$); this is inferred from $L_y \leq 19$. 
For $L_y=7$, we found roughly 
$\Nbas(L_x,7) \sim 3.5 {L_x}^{1/2} 2^{L_x}$, 
for $L_x$ up to 13.


\begin{thebibliography}{99}
\bibitem{FN-smalldroplets}
The smallest one with nonzero energies has 3 holes
inside a $3\sqrt 2 \times 3\sqrt 2$ square: the energies
are $\pm 2$ and (three-fold degenerate) $0$.
\bibitem{FN-stripevsdroplet}
Furthermore the two-stripe states had a lower energy, 
for the same particle number $M$. For a deeper discussion
of this issue, see the beginning of Sec.~\ref{sec-stripearray}.
\bibitem{FN-signrule}
In the case $L_y\equiv 3 \pmod{4}$, as
in Fig.~\ref{fig-stripe}, a simpler argument
is possible, which depends only on
the numbering convention
of Sec.~\ref{sec-diagstripe}.
A fermion sign can be introduced by a hop
only when a hop takes a particle across the $y$ boundary,
so the ordering of the $(L_y-1)/2$ particles in its column
undergoes a cyclic permutation with sign $\sigma_{\rm col}$
as defined in Eq.~(\ref{eq-sigmacol}).  This
is unity when $L_y\equiv 3 \pmod {4}$ so 
in Eq.~(\ref{eq-Hn}) $s(n,m)=1$ for every matrix element, 
i.e. the boson and fermion Hamiltonians are identical.
\bibitem{FN-Tch}
A generalization
to higher spins is straightforward if the 
stripe step can take more than two values. See
Ref.~\onlinecite{Tch} for a mapping of a
strongly anisotropic $t-J$ model to a spin-1 chain.
\bibitem {FN-1Dring} This description has glossed over the complications
that our one-dimensional wavefunctions are manifestly multi-valued. 
We must imagine that the many-body wavefunction of the 1D system 
jumps to another 
``Riemann sheet'' every time a 1D fermion hops across the boundaries. 
This is familiar from the theory of anyons.
\bibitem {FN-Lygauge}
The last term here does not yield the $\sigma_{\rm col}^h$  factor
described above, but it is gauge-equivalent to that. It gives
the correct factor for the $L_y$ iterated shifts by $(0,2)$
required to return to the original stripe configuration, 
which is $\sigma_{\rm col}^{L_x L_y}
= \sigma_{\rm col}^{L_x}$ -- recall $L_y$ must be odd.
\bibitem{FN-bosonLx}
For $(L_x,0)\times(0,7)$ and $(L_x,2)\times(0,7)$ systems
the boson and fermion energies are the same. For
$(L_x,1)\times(0,7)$ this is not true, and here we plot the
boson energies. Note that this does not 
invalidate the result (that boson and
fermion one-hole-with-a-stripe problems have
the same energy spectra) checked in Sec.~\ref{sec-onehole}
because it was for rectangular systems.
\bibitem{FN-holestep}
Snapshots of the high-weight configurations in the ground-state wavefunction
do commonly show the hole combined with an 
up-step as in Fig.~\ref{fig-45hole}, but we did not study this quantitatively.
\bibitem{FN-close}
This is not true for systems
with $L_x$ close to $L_y$. For example,
for $6\times 6$ with $M=12$, the lowest boson energy is 
$-7.98357$ and the fermion energy is $-7.92667$.
There we do not necessarily have two parallel stripes.
\bibitem{FN-reducedmass}
Alternatively, the same two-stripe problem can be
solved using relative coordinates (center-of-mass frame).
It appears as one particle with reduced mass $m_r=m^*/2$ 
confined to a box $(-L_y,+L_y)$. The energy is 
$E={\hbar^2 \pi^2}/{2m_r {L_y}^2}$, which agrees with 
Eq.~(\ref{eq-box}).
\bibitem{FN-phi3}
So strictly
speaking, we should write something like $\phi_{3b,f}^{L_x}$
to emphasize $\phi_3$'s dependence on $L_x$ and the boson
and fermion statistics, but when we do not explicitly study
these dependencies, we neglect them in $\phi_3$.
\bibitem{FN-lattices}
These are square lattices $(a,-b)(b,a)$ with $a=5,6$ and $b=0,1,2$,
and rectangular lattices $5\times 6$, $5\times 7$,
and $6\times 7$.
\end{thebibliography}
\end{document}